\documentclass[aip,pof,graphicx]{revtex4-2}
\usepackage{amssymb}
\usepackage{amssymb}
\RequirePackage{fix-cm}
\usepackage{graphicx}
\usepackage{graphicx}
\usepackage{epstopdf}
\usepackage{tikz}
\usepackage{xcolor}

\begin{document}

%\preprint{}

\title{An analytic \textcolor{black}{probability density function} for partially premixed flames with detailed chemistry} %Title of paper



\author{M. Pfitzner}
\email{michael.pfitzner@unibw.de}
\affiliation{Thermodynamics institute LRT-10, Bundeswehr University Munich, Werner-Heisenberg-Weg 39, 85579 Neubiberg, Germany.}
%\thanks{}
\author{P. Breda}
\affiliation{Thermodynamics institute LRT-10, Bundeswehr University Munich, Werner-Heisenberg-Weg 39, 85579 Neubiberg, Germany.}%Lines break automatically or can be forced with \\

\date{\today}

\begin{abstract}
Laminar premixed flame profiles of methane/air free flames and strained flames at different fuel/air ratios and strain rates are analysed using detailed chemistry with Lewis numbers equal to one. It is shown that the detailed chemistry flame profiles of progress variables CO2+CO and H2O+H2 in canonically stretched coordinates can be fitted accurately by a slight generalization of recently proposed analytical presumed flame profiles over a wide range of fuel/air ratios through adaptation of a single model parameter. Strained flame profiles can be reproduced using an additional linear coordinate transformation, emulating the compression of the preheat zone by strain as predicted by premixed flame theory. The model parameter can alternatively be determined using only the laminar flame speeds and the fully burnt temperatures from the laminar flame calculations. The stretch factor of the coordinate transformation is proportional to cp/lambda, which drops by factor up to 4 across the laminar flame. It is shown how the non-constant cp/lambda modifies the laminar flame \textcolor{black}{probability density function (pdf)}, a polynomial fit to cp/lambda as function of progress variable allows analytical results for the laminar flame pdf, the mean value of progress variable and of the reaction source term. An analytic pdf for partially premixed flames is proposed based on Bayes theorem as a combination of a beta pdf for the mixture fraction and the laminar flame pdf's evaluated at the respective fuel/air ratio. 
\end{abstract}


\maketitle %\maketitle must follow title, authors, abstract and \pacs

\section{Introduction}
In many technical burners featuring the propagation of thin flame fronts, the mixture fraction field is not completely homogeneous.Then flame fronts propagate through regions of varying fuel/air ratio but the thermal flame thickness is usually much smaller than the inverse gradient of mixture fraction. The reason is that small scale fluctuations of passive scalars like mixture fraction die out very quickly due to diffusion, while the gradient of progress variable within a propagating flame front is maintained by the interaction of diffusion and chemical reaction. 

In computational fluid dynamics (CFD) simulations of such flames using Reynolds-averaged Navier-Stokes (RANS) or Large Eddy Simulations (LES), the size of computational cells is usually too big to fully resolve the laminar flame structure embedded in the turbulent flow field. The ratio of flame thickness to cell size decreases further at elevated pressures due to the drop of diffusivities and heat conductivity. Thus subgrid combustion models are required for RANS and LES simulations of most technical burners.

The propagating flame fronts are folded and stretched by the turbulent flow field. Experimental observations \cite{driscoll2008turbulent} and Direct Numerical Simulation (DNS) results \cite{luca2019statistics} indicate that particularly at lower levels of turbulence intensity $u'/s_L$, the subgrid fuel consumption rate is increased proportionally to the amount of wrinkling of the reaction layer. Additional effects such as flame stretch, flame curvature and a thickening of the reaction layer through small scale turbulent eddies modify this proportionality only moderately. Even at quite large Karlovitz numbers, the inner structure of the reaction layer of hydrocarbon-air flames appears to remain largely intact \cite{nilsson2018structures}. 

In many turbulent combustion models for premixed flames, a single normalized reaction progress variable $c$ is invoked, which is zero (one) in the fully unburnt (burnt) regions. $c$ might be defined from the normalized temperature rise or by some combination of educt or product species. The local chemical state is well characterized by such a single progress variable as long as the inner flame structure is not strongly modified by turbulent mixing.

Premixed laminar flame profiles can be tabulated either as freely propagating flames or in a counterflow setting, where strained flames can be investigated. Once a monotonously rising progress variable is chosen, all other quantities can be calculated from such tables. (Quasi-) DNS calculations have been performed, where detailed chemistry was replaced by such flame generated manifold (FGM) tables of freely propagating premixed flames \cite{proch2017flame} while all turbulent eddies and the folding of the flame front was fully resolved.

Many turbulent premixed flame combustion subgrid models for LES and RANS simulations have been developed in the past. The artificially thickened flame  (ATF) model \cite{colin2000thickened} makes the flame front resolvable on LES grids by increasing the diffusion coefficient and the heat conductivity while reducing the reaction term such that the local laminar flame propagation speed remains unchanged. The effect of non resolved subgrid flame wrinkling is taken into account by an empirical efficiency function. 

Some models assume the existence of an infinitely thin flame front propagating at a turbulent flame speed $s_T$, which is provided through a separate model. Examples are the G equation level-set approach \cite{pitsch2002large} and subgrid flame surface density (FSD) models. In the latter, the sum of molecular diffusion term and chemical reaction source term of the $c$ transport equation is replaced by $\rho_u \left<s_c\right> \Sigma_f$ with flame surface density $\Sigma_f$ and a surface averaged flame speed $\left< s_c \right>$. $\Sigma_f$ is either determined by a transport equation \cite{poi05} or approximated as $\Sigma_f= \Xi \overline{\mid\nabla c\mid}$, evoking algebraic models for the wrinkling factor $\Xi$  \cite{ma2013posteriori} and often replacing $\overline{\mid\nabla c\mid}$ by $\mid\nabla\tilde{c}\mid$. Models of this type change the mathematical character of the  progress variable transport equation, preventing a recovery of the laminar flame front structure in the DNS limit.

In the framework of LES, the filtered laminar flame model pretabulates the chemical source term \cite{fiorina2010filtered}, which is filtered from a flat laminar flame on a one-dimensional (1D) grid. The filter size for the tabulation is either chosen equal to the LES grid size $\Delta$ or even larger than $\Delta$ to avoid numerical oscillations. In the latter case, the chemical source term is smeared out over a larger number of cells in physical space. Incorporation of effects of subgrid flame folding, of flame stretch and flame thickening requires empirical modifications. Attempts to derive more accurate expressions for $\overline{\omega(c)}$ for a underresolved flat flame in LES cells by 1-D approximate deconvolution have been reported in \cite{domingo2015large}. The latter method is not easily generalizable to more the one spatial dimension and appears to require quite fine LES resolution.

In contrast to models developed specifically for premixed flames, pdf methods are formally applicable to all combustion regimes and to an arbitrary number of spatial dimensions. The structure of the subgrid flame front is reflected in the shape of the pdf and this has not always been appreciated enough in the past. While pdf methods have been very successful in the modelling of non-premixed flames, the straightforward application of those methods to estimate the filtered source term in the $c$ transport equation of premixed flames can yield inaccurate results.

The beta pdf is a good model pdf for pure diffusion processes and it represents the standard subgrid pdf for mixture fraction in non-premixed flames. As a progress variable pdf in premixed flames, it has been shown to overestimate the mean reaction term  for large $c$ variance near to the thin flame limit \cite{bray2006finite}. DNS analyses \cite{knudsen2010analysis} showed good agreement between the DNS source term filtered to a LES grid and the beta pdf value (with mean and variance in the beta pdf evaluated from the DNS) only for small ratios of ${\Delta_{LES} / \Delta_{DNS}} < 5$, i.e. $\Delta_{LES} < \delta_{th}$ where $\delta_{th}$ is the thermal flame thickness. 

Presumed premixed flame pdfs derived from filtering numerically generated 1D laminar flame profiles have been proposed by \cite{bray2006finite}$^,$\cite{domingo2005dns}$^,$ \cite{salehi2010presumed}. These authors invoked empirical cutoffs near $c=0,1$ to avoid numerical divergence of their integrals when calculating the normalization condition of the pdf and mean and variance of the progress variable.  Domingo  et al.\cite{domingo2005dns} report that such laminar flame pdfs with ad hoc choice of the integration limits $c^-,c^+$ delivered negative weights of the delta functions for small values of $c$ variance. The authors used a beta pdf in those cases. Salehi et al.\cite{salehi2010presumed}$^,$\cite{salehi2013modified} presented a modified laminar flame pdf allowing its application to the whole range of $c$ variances. The application of the (stochastic) linear eddy model to numerically derived 1D laminar flames\cite{jin2008conditional} and evaluation of DNS data\cite{tsui2014linear} showed pdf's with smeared out cutoffs in the regions towards $c=0,1$.

Subgrid progress variable pdf's were evaluated from DNS data by Moureau et al.\cite{moureau2011large} and Lapointe and Blanquart\cite{lapointe2017priori}. To gain a good agreement of the level of the DNS pdfs with 1D laminar flame pdfs, the latter were filtered using an effective filter width $\Delta'<\Delta$, which was calculated from the condition that the mean and variance evaluated with the 1D pdf agreed with those evaluated from the filtered DNS.  

Analytical laminar flame shapes and pdfs can be evaluated when using appropriate surrogate reaction source terms instead of the complex Arrhenius ones. A simple linear source term\cite{ferziger1993simplified} was proposed in the 1990's, recently a more accurate approximation to the Arrhenius single-step chemistry source term was introduced\cite{pfitzner2020pdf}. Both surrogate source terms generate analytical flame profiles and laminar flame pdfs. A simple two-dimensiontal sinusoidal flame folding model\cite{pfitzner2020pdf} revealed that flame wrinkling increases the level of the pdf in the reactive region while it is smeared out near the cutoffs near $c=0,1$. The increased level of the pdf in the reactive $c$ region could be emulated through a laminar flame pdf evaluated with a reduced filter size $\Delta'=\Delta/\Xi$ where $\Xi$ is the geometrical wrinkling factor of the folded 
reaction layer. This shows that the use of a (smaller) effective filter width in the work of Moureau and
Lapointe/Blanquart effectively mimics the effect of subgrid flame wrinkling.

The goal of the present paper is to show that using an appropriately (canonically) stretched coordinate, premixed laminar methane-air flame profiles generated with detailed chemistry can be represented very accurately through a analytically defined, invertable flame profiles over a large range of fuel/air ratios and strain rates, providing analytical expressions for mean values of progress variable, reaction source term and laminar flame pdf. 

The paper is structured as follows: first we present the progress variable transport equation and recall some ingredients of single-step Arrhenius chemistry and the canonical transformation of the spatial coordinate. After introduction of the analytical presumed flame profile and pdf we describe the calculation of the premixed laminar flame profiles using detailed chemistry. We discuss the process of fitting the model parameter to flame profile and laminar flame speed and derive correlations for use in actual simulations. We then propose a 2-D analytical presumed pdf $p(Z,c)$ for partially premixed flames based on Bayes' theorem and give some conclusions.

\section{Transport equation of progress variable}

In flames where thin premixed flame fronts are propagating through a (potentially inhomogeneous) mixture, it is common to use a single reaction progress variable, which is often chosen as $c=(T-T_u)/(T_b-T_u)$ in the homogeneous mixtures at constant pressure. $T_u,T_b$ are the unburnt and fully burnt temperatures, respectively. Alternatively, a suitable combination of concentrations of chemical species can be chosen, which may be more suitable in the inhomogeneous case due to the dependence of $T_b$ on fuel/air ratio. The one dimensional $c$ transport equation is given by \cite{poi05}
\begin{equation} 
	\rho\frac{\partial c}{\partial t}+\rho u \frac{\partial c}{\partial x}=\frac{\partial}{\partial x}\left(\frac{\lambda}{c_p} \frac{\partial c}{\partial x}\right)-\frac{\dot\omega_F}{Y_{F}}              
	\label{eq:ctrans}
\end{equation}
where $\rho$, $u$, $c$ are density, velocity and progress variable, and $\lambda$, $c_p$ are the heat conductivity and specific heat at constant pressure. For Arrhenius chemistry and Lewis number $Le=1$, the chemical source term can be written as \cite{poi05}
\begin{equation} 
	\frac{\dot{\omega}_F}{Y_{F}}=B_1 T^{\beta_1} e^{-\frac{\beta}{\alpha}}\rho(1-c)exp\left(-\frac{\beta(1-c)}{1-\alpha(1-c)}\right)
	\label{eq:omsource}
\end{equation}
where $\alpha=\frac{T_b-T_u}{T_b}$ represents the normalized temperature raise and $\beta=\alpha T_{a}/T_b$ is a measure of the activation temperature $T_{a}$. The temperature exponent $\beta_1$ in 
eq.(\ref{eq:omsource}) is usually taken as $\beta_1=0$ or $\beta_1=1$. For steady-state conditions, the continuity equation requires 
$\rho u=\rho_u s_L$ and we have $\rho \sim 1/T \sim 1/\left( 1-\alpha(1-c)\right)$ for constant pressure combustion. 

For a stationary flame eq.(\ref{eq:ctrans}) yields
\begin{equation} 
	\rho_u s_L \frac{\partial c}{\partial x}-\frac{\partial}{\partial x}\left(\frac{\lambda}{c_p} \frac{\partial c}{\partial x}\right)=\omega_x(c)              
	\label{eq:ctransscal}
\end{equation}
with $\omega_x(c)=-\frac{\dot\omega_F}{Y_{F}}$ and taking into account the continuity equation $\rho u= \rho_u s_L$. 
Rescaling the $x$ coordinate according to $d\xi=\rho_us_Lc_p/\lambda dx$ as
\begin{equation}
	\xi-\xi_0 = \rho_u s_L \int_{x_0}^x (c_p / \lambda) dx'
	\label{eq:xitrans}
\end{equation}
transforms eq.(\ref{eq:ctrans}) into canonical form:
\begin{equation} 
	\frac{\partial c}{\partial \xi}-\frac{\partial^2 c}{\partial \xi^2}=\omega(c)              
	\label{eq:cstrans}
\end{equation}
with
\begin{equation} 
	\omega(c)=\Lambda \left(1-\alpha(1-c)\right)^{\beta_1-1} (1-c)exp\left(-\frac{\beta(1-c)}{1-\alpha(1-c)}\right)=\left(\frac{\lambda}{c_p}\right)\frac{\omega_x(c)}{(\rho_u s_L)^2}              
	\label{eq:omstrans}
\end{equation}
The prefactor $\Lambda$ in eq.(\ref{eq:omstrans}) represents the Eigenvalue of the transport equation which guarantees that the boundary conditions $c = 0$ for $\xi\rightarrow -\infty$ and  $c = 1$ for $\xi\rightarrow +\infty$ are fulfilled. 

If the progress variable is not normalized, i.e. $c = C$ for $\xi\rightarrow +\infty$ with $C \neq 1$, the source term is just multiplied by $C$. Note that $c_p/\lambda$ is not constant in eq.(\ref{eq:xitrans}) for flame profiles calculated with detailed chemistry and realistic transport properties, leading to a nonlinear stretch transformation. 

\section{Laminar flame pdf's}
A probability density function $p(c)$ allows the evaluation of cell averages of arbitrary quantities $z(c)$ through $\overline{z(c)}=\int z(c) p(c) dc$.  In classical Bray-Moss-Libby (BML) theory the pdf is assumed to take the form
\begin{equation} 
	p_{BML}(c)=A \delta(c) + B \delta(1-c) + \gamma(c)
	\label{eq:BML}
\end{equation}
with $\gamma(c) \ll 1$. For $\gamma(c)\rightarrow 0$ one obtains  $A \sim (1-\overline{c})$ and $B \sim \overline{c}$. Since the chemical source term $\omega(c)$ vanishes at $c=0,1$, its mean cannot be evaluated from $p_{BML}(c)$ with $\gamma(c)$ set to zero. 
For a given $c(\xi)$ profile, the 1D laminar flame pdf $p(c)$ is given by \cite{bray2006finite}$^,$\cite{moureau2011large}$^,$\cite{pfitzner2020pdf}:
\begin{equation} 
	p(c) = \frac{1}{N}\frac{1}{dc/d\xi}H(c-c^-)H(c^+-c)
	\label{eq:pdfdef}
\end{equation}
where $H(x)$ is the Heaviside function and $c^-=c(\xi^-),c^+=c(\xi^+)$ with $\xi^-,\xi^+$ denoting the left and right boundaries of the filter interval. The denominator $N$ guarantees the correct normalisation of $p(c)$:
\begin{equation} 
	\int_0^1p(c)dc=\frac{1}{N}\int_{c^-}^{c^+}\frac{1}{dc/d\xi}dc=\frac{1}{N}\int_{\xi^-}^{\xi^+} d\xi=\frac{\xi^+-\xi^-}{N}=1
	\label{eq:pdfnor}
\end{equation}
yielding $N=\xi^+-\xi^-=\Delta_\xi$, which is true for any 1D laminar flame pdf. For constant $c_p/\lambda$, $N$ is directly proportional to the filter width $\Delta_x$ in $x$ space. 
The mean of any variable $z(c)$ evaluates as:
\begin{equation} 
	\overline{z(c)}=\int_0^1z(c)p(c)dc=\frac{1}{N} \int_{c^-}^{c^+}\frac{z(c)}{dc/d\xi} dc=\frac{1}{(\xi^+-\xi^-)} \int_{\xi^-}^{\xi^+}z(\xi) d\xi
	\label{eq:meanwc}
\end{equation}
proving the validity of these results, since the last term represents the correct 1D spatial mean of $z$. 

\section{Evaluation of filtered source term}
Analytical evaluation of $\overline{\omega(c)}$ for a given $c(\xi)$ is always possible for a given analytical flame profile $c(\xi)$, using eq.'s (\ref{eq:cstrans},\ref{eq:meanwc}):
\begin{equation}
	\overline{\omega(c)}=\frac{1}{(\xi^+-\xi^-)} \int_{\xi^-}^{\xi^+}\left(
	-\frac{\partial^2 c}{\partial \xi^2}+\frac{\partial c}{\partial \xi}\right) d\xi=\frac{1}{(\xi^+-\xi^-)}\left[
	-\frac{\partial c(\xi)}{\partial \xi}+c(\xi)\right]^{\xi^+}_{\xi^-}
	\label{eq:ommean}
\end{equation}
An analytic inversion $\xi(c)$ of $c(\xi)$ allows to convert the last term on the RHS into an expression depending on $c$ instead of $\xi$. The lower and upper boundaries $\xi^-,\xi^+$ then translate into the lower and upper boundaries of the pdf, $c^-,c^+$.

The mean of the sum of laminar diffusion and reaction source terms used in many flame surface density models can be evaluated exactly for any 1D laminar flame pdf:
\begin{equation} 
	\overline{\frac{\partial^2 c}{\partial \xi^2}+\omega(c)}=\overline{\frac{\partial c}{\partial \xi}}=\frac{1}{(\xi^+-\xi^-)} \int_{c^-}^{c^+}\frac{\frac{\partial c}{\partial \xi}}{dc/d\xi}dc=\frac{1}{(\xi^+-\xi^-)}\int_{c^-}^{c^+}dc=\frac{(c^+-c^-)}{(\xi^+-\xi^-)}
	\label{eq:omdifmeanm}
\end{equation}

\section{Presumed flame profile}
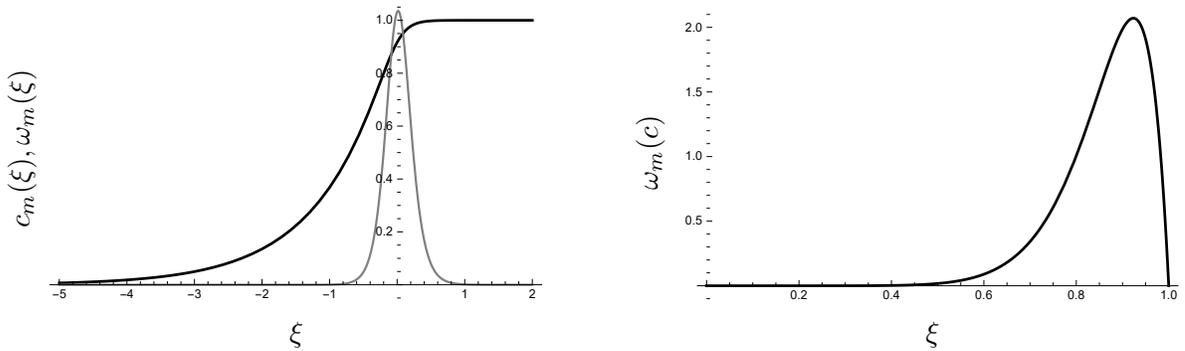
\begin{figure} [ht]
	\begin{minipage}[b]{.4\linewidth} % [b] => Ausrichtung an \caption
		\begin{tikzpicture}
			\node[] (Grafik) at (0,0) {\includegraphics[width=1\textwidth]{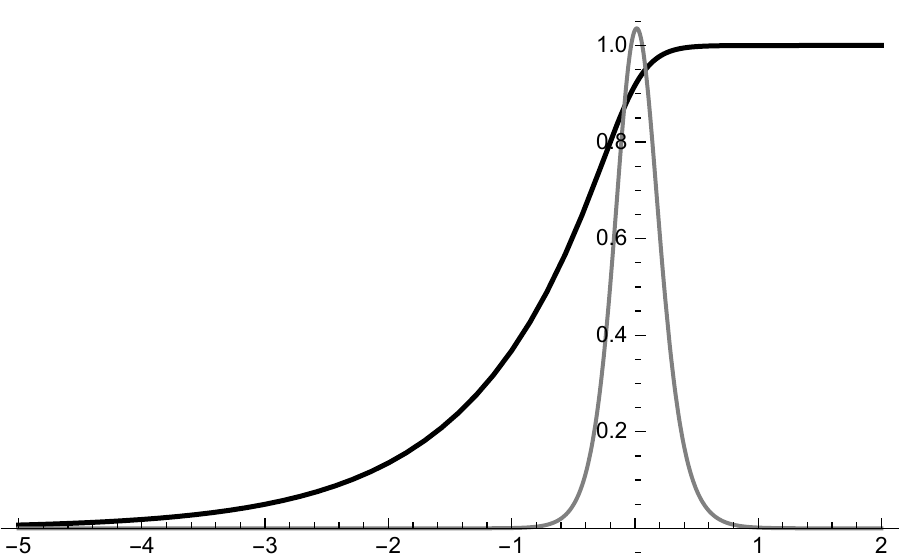}};
			\node[anchor=north,yshift=0pt,xshift=0] at (Grafik.south) {$\xi$};
			\node[rotate=90,anchor=south,xshift=0pt,yshift=-3] at (Grafik.west) {$c_m(\xi), \omega_m(\xi)$};
		\end{tikzpicture}
	\end{minipage}
	\hspace{.1\linewidth}% Abstand zwischen Bilder
	\begin{minipage}[b]{.4\linewidth} % [b] => Ausrichtung an \caption
		\begin{tikzpicture}
			\node[] (Grafik) at (0,0) { \includegraphics[width=\linewidth]{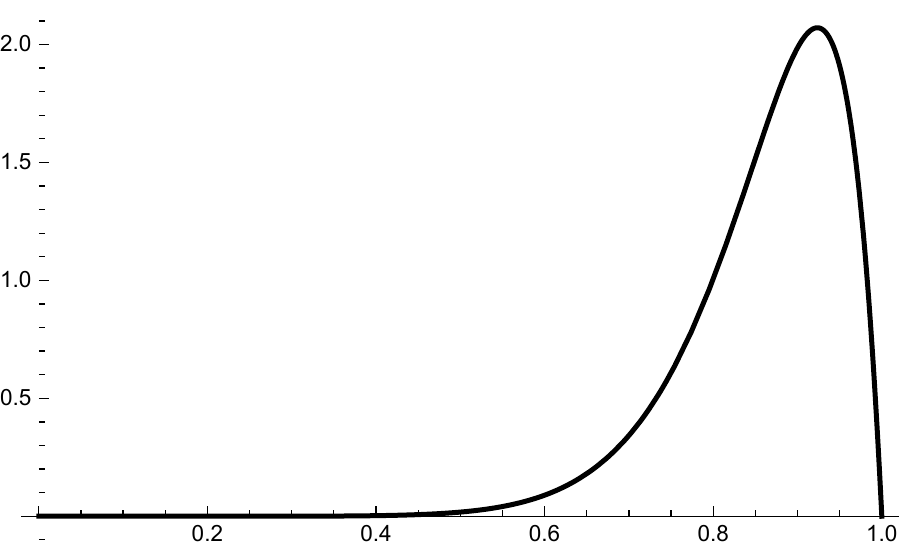}};
			\node[anchor=north,yshift=-0pt,xshift=0] at (Grafik.south) {$\xi$};
			\node[rotate=90,anchor=south,yshift=-0,xshift=0pt] at (Grafik.west) {$\omega_m(c)$};
		\end{tikzpicture} 
	\end{minipage}
	\caption{Left: Presumed flame profile $c_m(\xi)$ (black) and source term $\frac{1}{2} \cdot \omega_m(\xi)$ (gray); right: source term $\omega_m(c)$}
	\label{fig:cromm8}
\end{figure}

In a recent contribution \cite{pfitzner2020pdf}, an analytic invertable progress variable profile $c_m(\xi)$ was derived which also provides an integrable source  term $\omega_m(c)$ with a parameter $m$ which can be adapted to different activation energy situations. This profile is generalized here slightly with the introduction of two additional constants $C,a$:
\begin{equation} 
	c_m(\xi)=\frac{C}{[1+\exp(-a \cdot m \cdot \xi)]^{1/m}}              
	\label{eq:cmgen}
\end{equation}
The new parameter $C$ represents the maximum of the progress variable in the fully burnt state. Parameter $a$ provides the possibility to rescale the $\xi$ coordinate by a spatially constant factor which will be useful in the representation of flame profiles of strained flames. Fig.(\ref{fig:cromm8}) shows the presumed flame profile and the corresponding source term $\omega(\xi)$ (scaled by a factor of 0.5 for clarity) for $m=8$. Also shown is the source term $\omega(c)$ in $c$ coordinates.

$c_m(\xi)$ can be inverted as
\begin{equation}
\xi_m(c)=-\frac{\log \left(\left(\frac{c}{C}\right)^{-m}-1\right)}{a \cdot m}
\end{equation}
The spatial derivative of $c(\xi)$, expressed in $c$ itself is given by
\begin{equation}
	dc/d\xi= a \cdot  c (1 - (c/C)^m)
	\label{eq:dcdxi}
\end{equation}
yielding a thermal flame thickness (which is the inverse of the maximum derivative) of
\begin{equation} 
	\delta_{th}=\frac{1}{\left( dc/d\xi \right)_{max}}=\frac{(m+1)^{\frac{m+1}{m}}}{a \cdot C \cdot m}
	\label{eq:deltafn}
\end{equation}
For constant stretch factor (i.e. $c_p/\lambda$ = const.), the laminar flame pdf evaluates as
\begin{equation}
	p(c)=\frac{1}{\Delta_\xi}\frac{1}{a \cdot  c (1 - (c/C)^m)}H(c-c^-)H(c^+-c)
\end{equation}
with $\Delta_\xi=\left(\rho_us_Lc_p/\lambda\right) \cdot \Delta_x$.
The chemical source term evaluates as
\begin{equation}
	\omega_m(c)=a\cdot c (1 - (c/C)^m) (1 - a (1 - (c/C)^m (1 + m)))
	\label{eq:omegaCa}
\end{equation}
with mean value
\begin{equation}
\overline{\omega_m(c)}=\int_0^1 \omega_m(c) p(c)dc=\frac{1}{N} \int_{c^-}^{c^+}\frac{\omega_m(c)}{dc/d\xi} dc
\end{equation}
For constant $c_p/\lambda$ we obtain
\begin{equation}
	\overline{\omega_m(c)}=\frac{1}{\Delta_\xi} \int_{c^-}^{c^+}(1 - a (1 - (m+1)(c/C)^m) dc=
	\frac{1}{\Delta_\xi}\left[ c (1-a(1 - (c/C)^m))\right]_{c^-}^{c^+}
	\label{eq:intomegaCa}
\end{equation}
For a filter interval in physical space with boundaries $[x^-,x^+]$, transforming into $[\xi^-,\xi^+]$, the limits of the integration evaluate as $c^-=c_m(\xi^-)$ and $c^+=c_m(\xi^+)$.
In the case $a=1, C=1$, eq.(\ref{eq:intomegaCa}) reduces to\cite{pfitzner2020pdf} the simple result
\begin{equation}
	\overline{\omega_m(c)}=\frac{1}{\Delta_\xi} \left((c^+)^{m+1}-(c^-)^{m+1}\right)
	\label{eq:intomegaCa1}
\end{equation}

\section{Generation of flame profiles}
The methane-air laminar premixed flames are generated in this work using the software CANTERA \cite{Goodwin2018} v. 2.4. The Gas Research Institute (GRI) mech 3.0 \cite{Frenklach1995} detailed chemical mechanism with assumption of unity Lewis numbers was used in the calculations. The temperature of the unburned mixture is set to T$_u$\,=\,300\,K and combustion occurs at atmospheric conditions (p\,=\,1\,bar).

The freely-propagating flat flames are calculated over a range of equivalence ratios $\phi$ varying from 0.4 to 2.2, on a 20\,mm wide physical grid. Stretched premixed flames at equivalence ratios of $\phi=1.0$ and $\phi=0.6$ are calculated using a double premixed counter-flow flame configuration, where two identical, axially symmetric, premixed jets of fresh gases blow against each other. In these cases, the half-domain is 40\,mm wide. The flame strain \textit{K} varied from $K=10$ to $K=1000\,s^{-1}$ and was achieved by changing the inlet jet velocity \textit{U}. The value of \textit{K} is retrieved as the maximum of the local strain rates observed upstream, before the pre-heating zone. 

\begin{figure}[hbt!]
	\centering
	\includegraphics[width=1\textwidth]{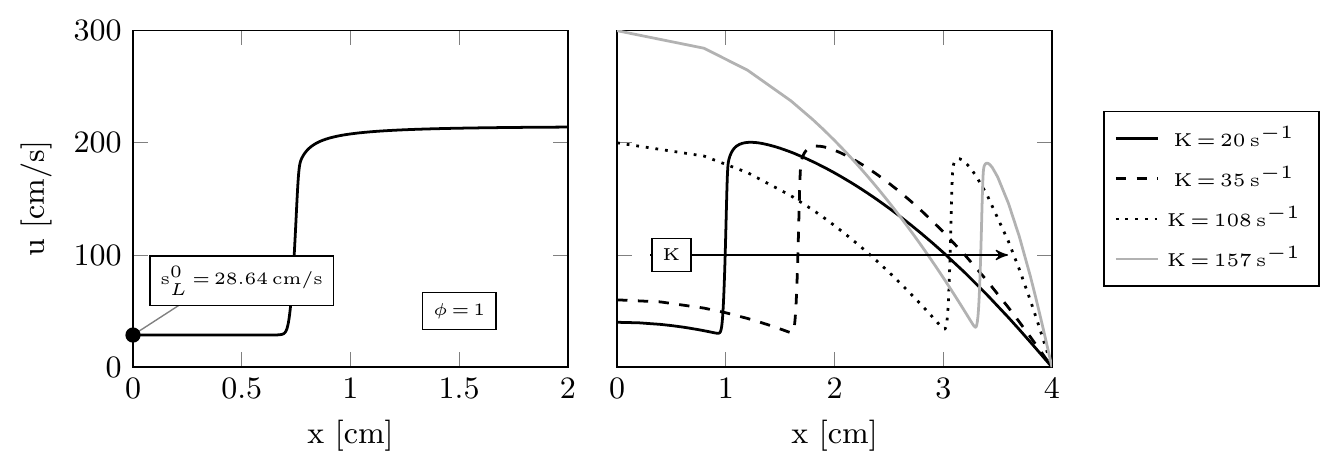}
	\caption{Left: laminar flame speed s$_L^0$\,=\,28.64\,cm/s for free flame at $\phi$\,=\,1. Right: stretched flames at different \textit{K} for $\phi$\,=\,1}
	\label{fig:sC}
\end{figure}

Fig.\,\ref{fig:sC} (left) reports the \textit{u} profile across the steady free flame at $\phi$\,=\,1, where the laminar flame speed s$_L^0$ is taken as the speed of the fresh gas retrieved at x\,=\,0. In this example, the flame propagation speed $s_L^0\,=\,28.64\,cm/s$. This value cannot be retrieved at the same location in the stretched flames since the velocity in x\,=\,0 is the imposed inlet velocity, as shown on the right plot of Fig.\,\ref{fig:sC}. A local minimum is seen before the pre-heating zone (steep-increase in \textit{u}) and its position moves towards the stagnation point (x\,=\,40\,mm) by increasing the strain \textit{K}. The velocity drops to zero at the stagnation point. 

\noindent In order to quantify the flame speed uniquely for both configurations, the flame consumption speed is used in this work. It is defined as the integral of the heat release rate across the flame brush

\begin{equation}
	s_c = \frac{1}{(T_b-T_u)\,\rho_u} \int \frac{\dot{Q}}{c_p} \,dx
	\label{eq:sC}
\end{equation}

\noindent with $\rho_u$ being the density of the unburned mixture, T$_b$ the temperature of the burned mixture, $c_p$ the mixture heat capacity at constant pressure and $\dot{Q}$ the total heat release rate. 

\begin{figure}[hbt!]
	\centering
	\includegraphics[width=0.8\textwidth]{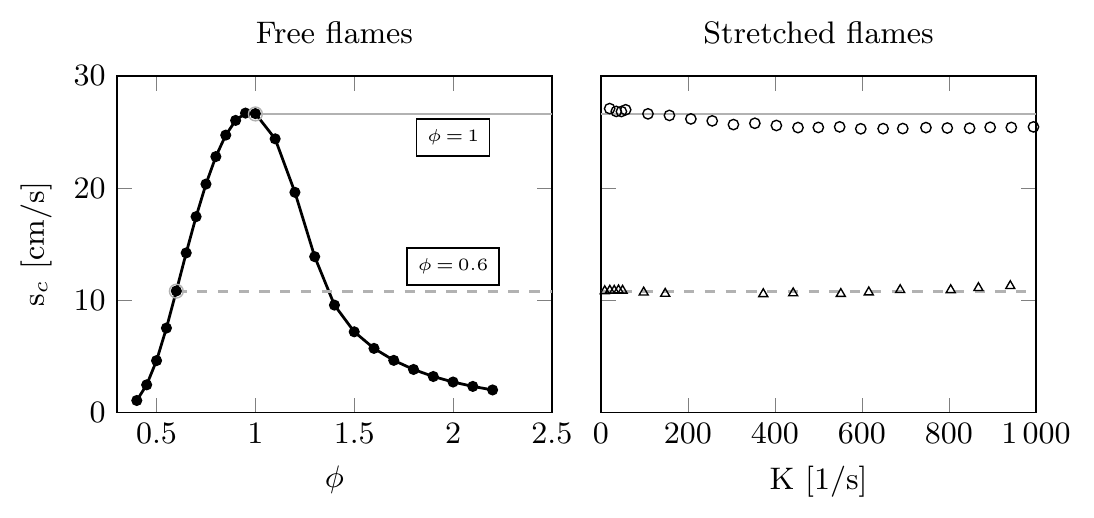}
	\caption{Left: Free flames consumption speed s$_c$ over the investigated $\phi$ range. Right: s$_c$ for stretched premixed flames at $\phi$\,=\,0.6 and 1, over an increasing range of \textit{K}}
	\label{fig:phiRange}
\end{figure}

The evolution of \textit{s}$_c$ over the investigated range of $\phi$ is shown for the free flames in Fig.\,\ref{fig:phiRange}. The plot on the right shows how the laminar flame is affected by strain for $\phi$\,=\,1 and 0.6. At higher strains the consumption speed is slightly reduced, but it tends to the value calculated for the free flames by decreasing \textit{K}, as expected.

Each stored flame profile contains additional scalars calculated at the grid points. The species progress variables (PV) used in this work are defined as a linear combination of specific mole numbers. The specific mole number of species \textit{k} is defined as $\frac{Y_k}{W_k}$, with mass fraction Y$_k$ and species molecular weight W$_k$. 

A preliminary investigation showed that $PV_1=CO_2+CO$ representing carbon chemistry and $PV_2=H_2O+H_2$ representing hydrogen chemistry appeared suitable progress variables due to a monotonous rise and positive source terms, which are calculated as $\omega_s=\sum_{i=1}^{n_r} \left( \frac{\dot{\omega_i}\,W}{\rho} \right)$, with $n_r$ being the number species used to define the PV (therefore two in this work), $\dot{\omega_i}$ the net production rate of the species \textit{i} and W the mean molecular weight. The values of $\lambda/c_p$ required in Eq.\,\ref{eq:xitrans} and the $\dot{Q}/c_p$ for Eq.\,\ref{eq:sC} were stored in addition in the tables.

\section{Analysis of GRI mech 3.0 progress variables}
In a first step we investigated whether the analytic $c_m(\xi)$ flame profiles, which were developed in the context of single-step Arrhenius chemistry, could be suitable to also represent progress variable profiles derived from flame calculations with detailed chemistry and transport. In addition to the above mentioned progress variables $PV_1,PV_2$, we investigated several other progress variables used in the literature (e.g. mass and mole fractions of $CO_2$, $CO_2+CO$, $H_2O$, $CO_2+CO+H_2O+H_2$, normalized temperature) for similarity to the single-step chemistry profiles. 

\begin{figure} [ht]
	\begin{minipage}[b]{.42\linewidth} % [b] => Ausrichtung an \caption
		\begin{tikzpicture}
			\node[] (Grafik) at (0,0) {\includegraphics[width=1\textwidth]{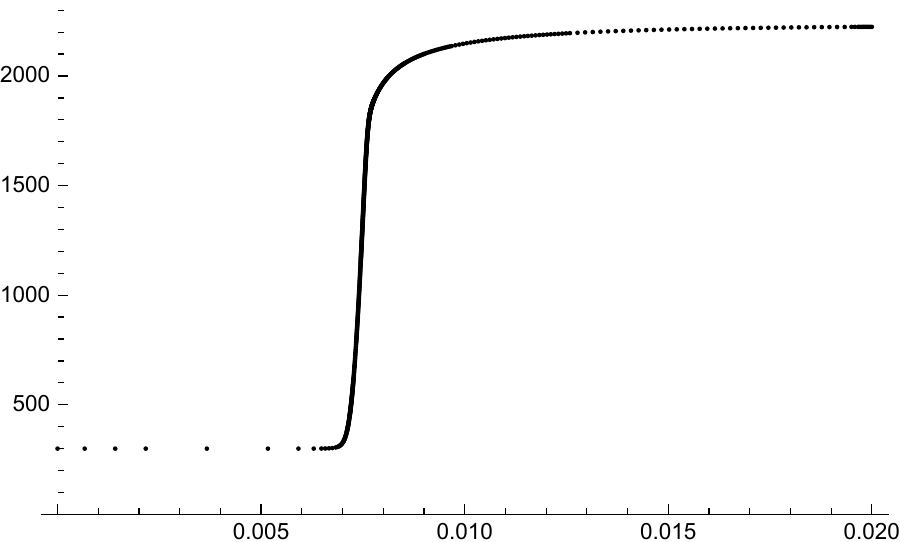}};
			\node[anchor=north,yshift=0pt,xshift=0] at (Grafik.south) {$x [m]$};
			\node[rotate=90,anchor=south,xshift=0pt,yshift=-3] at (Grafik.west) {$T [K]$};
		\end{tikzpicture}
	\end{minipage}
	\hspace{.1\linewidth}% Abstand zwischen Bilder
	\begin{minipage}[b]{.42\linewidth} % [b] => Ausrichtung an \caption
		\begin{tikzpicture}
			\node[] (Grafik) at (0,0) { \includegraphics[width=\linewidth]{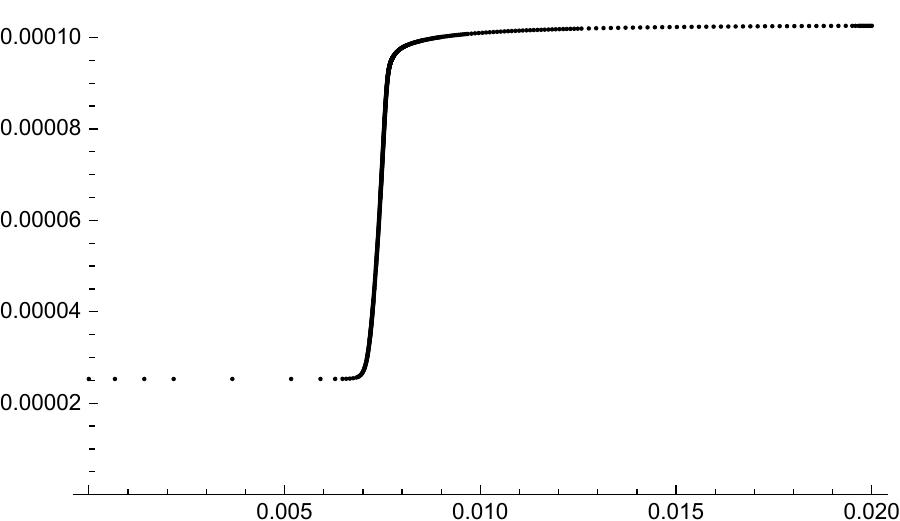}};
			\node[anchor=north,yshift=-0pt,xshift=0] at (Grafik.south) {$x [m]$};
			\node[rotate=90,anchor=south,yshift=-0,xshift=0pt] at (Grafik.west) {$\lambda/c_p$};
		\end{tikzpicture} 
	\end{minipage}
	\caption{Temperature (left) and $\lambda/c_p$ (right) for $\phi=1.0$ free flame}
	\label{fig:Tlamcp}
\end{figure}
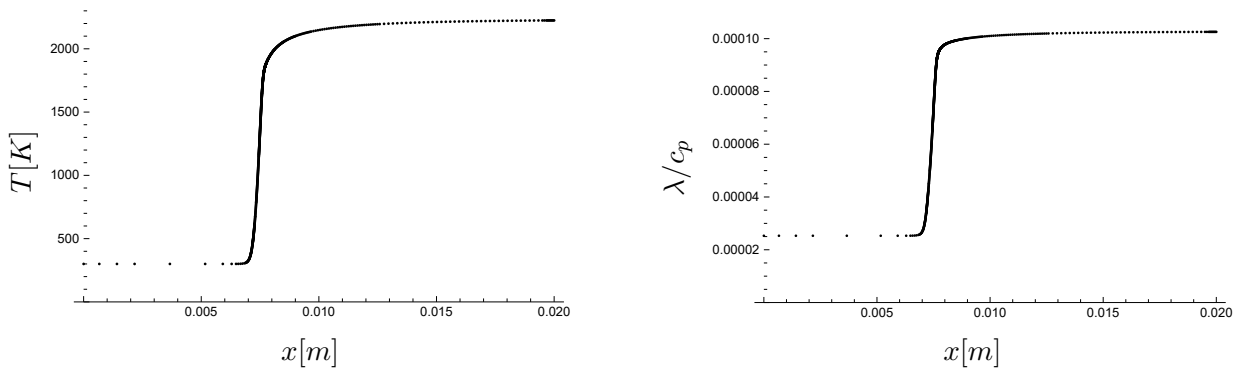

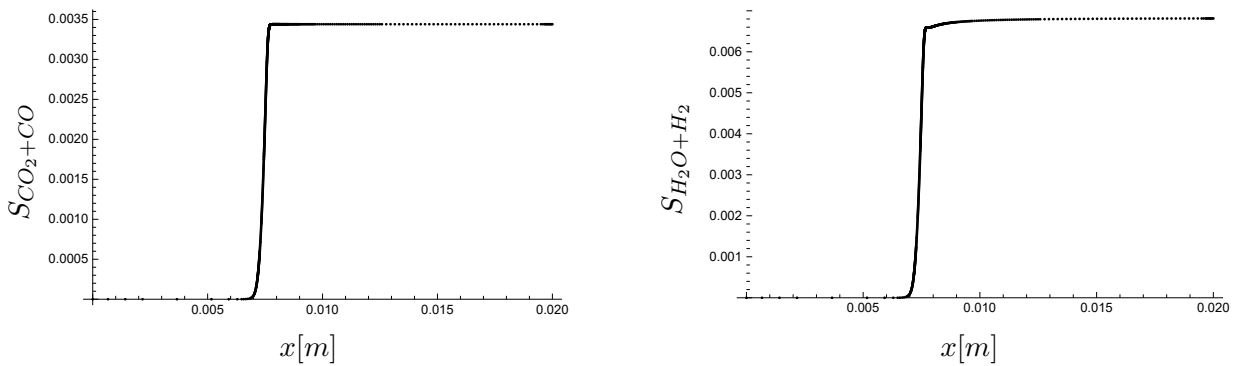
\begin{figure} [ht]
	\begin{minipage}[b]{.42\linewidth} % [b] => Ausrichtung an \caption
		\begin{tikzpicture}
			\node[] (Grafik) at (0,0) {\includegraphics[width=1\textwidth]{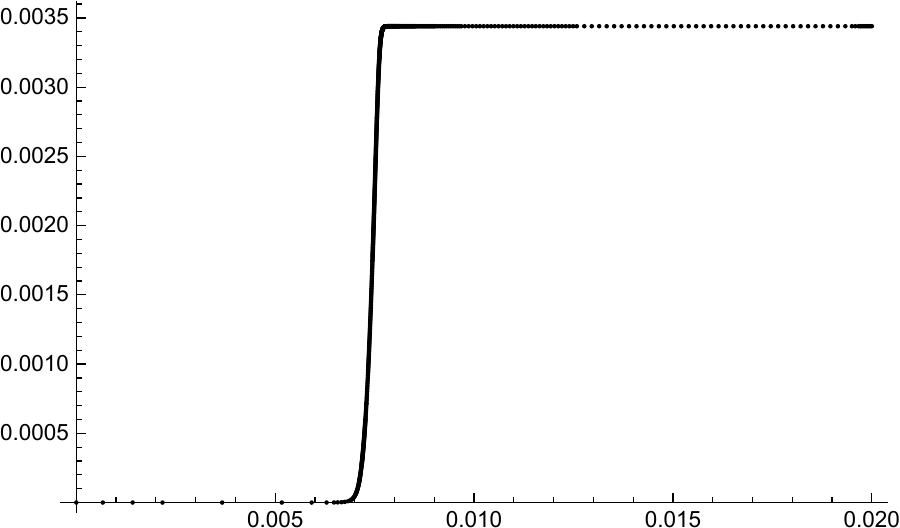}};
			\node[anchor=north,yshift=0pt,xshift=0] at (Grafik.south) {$x [m]$};
			\node[rotate=90,anchor=south,xshift=0pt,yshift=-3] at (Grafik.west) {$S_{CO_2+CO}$};
		\end{tikzpicture}
	\end{minipage}
	\hspace{.1\linewidth}% Abstand zwischen Bilder
	\begin{minipage}[b]{.42\linewidth} % [b] => Ausrichtung an \caption
		\begin{tikzpicture}
			\node[] (Grafik) at (0,0) { \includegraphics[width=\linewidth]{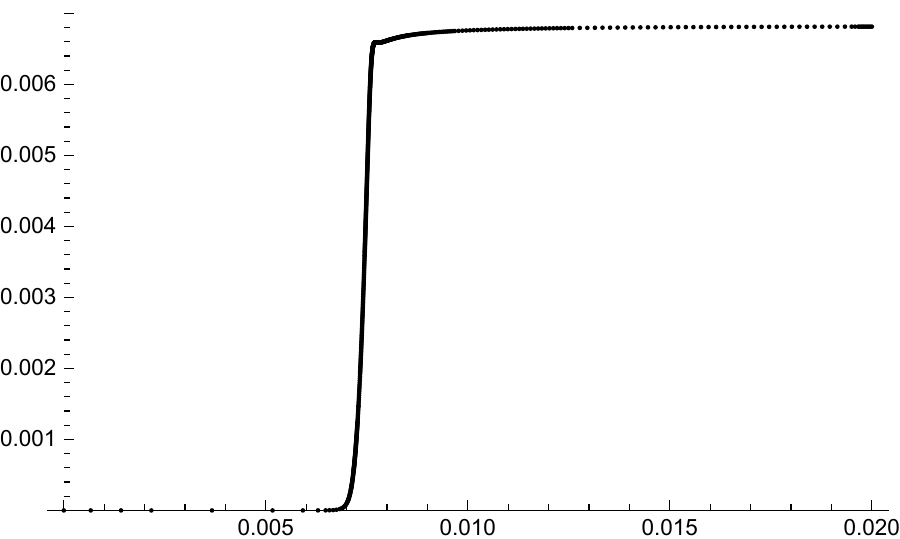}};
			\node[anchor=north,yshift=-0pt,xshift=0] at (Grafik.south) {$x [m]$};
			\node[rotate=90,anchor=south,yshift=-0,xshift=0pt] at (Grafik.west) {$S_{H_2O+H_2}$};
		\end{tikzpicture} 
	\end{minipage}
	\caption{Specific mole number of $CO_2+CO$ (left) and $H_2O+H_2$ (right) for $\phi=1.0$ free flame}
	\label{fig:CO2COH2OH2}
\end{figure}

Fig.(\ref{fig:Tlamcp}) shows the temperature and $\lambda/c_p$ distributions of a $\phi=1$ free flame in $x$ space. The temperature distribution features a diffusive preheat zone and a steep  reaction region which is followed by a long tail on the burnt side. The $\lambda/c_p$ plot indicates that a strongly nonlinear stretch will occur within the flame front.
We found that while most mass fraction and mole fraction combinations show a similar long tail on the burnt side like normalized temperature, which cannot directly  be reproduced by $c_m(\xi)$, the above mentioned combination of specific mole numbers (being the mass fraction divided by the corresponding molar weight) of $CO_2+CO$ (representing carbon oxidation) and $H_2O+H_2$ (representing hydrogen oxidation) yielded profiles looking very similar to one-step chemistry, see Fig.(\ref{fig:CO2COH2OH2}). While $CO_2+CO$ looks like a perfect match with one-step chemistry profiles, the $H_2O+H_2$ additionally features a long tail on the burnt side, which in this case is however restricted to values very near to $c=1$.

In the presentation of results, we will focus on analysis of the $CO_2+CO$ profiles. $H_2O+H_2$ yielded very similar results in the range $0 \leq c \leq C \approx 0.97$. We will also provide an analytical approximation of the $c$ profile in the tail region applicable to $H_2O+H_2$ but also to other progress variables like normalized temperature.

\section{Transformation of flame profiles}
\begin{figure} [ht]
	\begin{minipage}[b]{.42\linewidth} % [b] => Ausrichtung an \caption
		\begin{tikzpicture}
			\node[] (Grafik) at (0,0) {\includegraphics[width=1\textwidth]{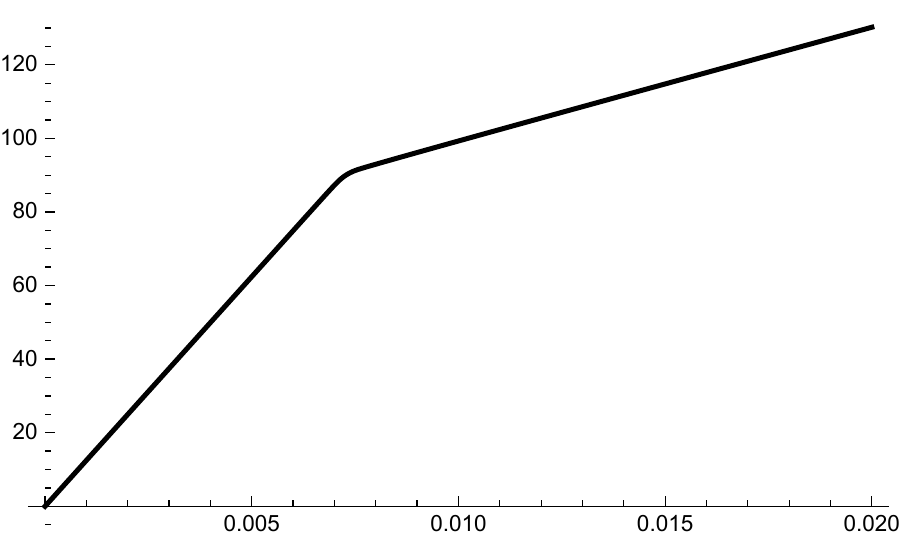}};
			\node[anchor=north,yshift=0pt,xshift=0] at (Grafik.south) {$x [m]$};
			\node[rotate=90,anchor=south,xshift=0pt,yshift=-3] at (Grafik.west) {$\xi(x)$};
		\end{tikzpicture}
	\end{minipage}
	\hspace{.1\linewidth}% Abstand zwischen Bilder
	\begin{minipage}[b]{.42\linewidth} % [b] => Ausrichtung an \caption
		\begin{tikzpicture}
			\node[] (Grafik) at (0,0) { \includegraphics[width=\linewidth]{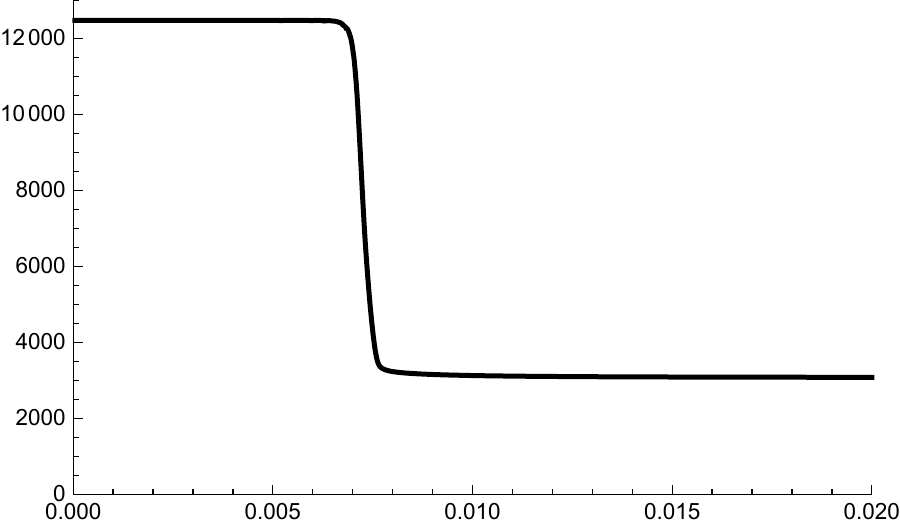}};
			\node[anchor=north,yshift=-0pt,xshift=0] at (Grafik.south) {$x [m]$};
			\node[rotate=90,anchor=south,yshift=-0,xshift=0pt] at (Grafik.west) {$d\xi/dx$};
		\end{tikzpicture} 
	\end{minipage}
	\caption{Transformed coordinate $\xi(x)$ (left) and its derivative $d\xi/dx$ (right); $\phi=1.0$}
	\label{fig:xidxix}
\end{figure}
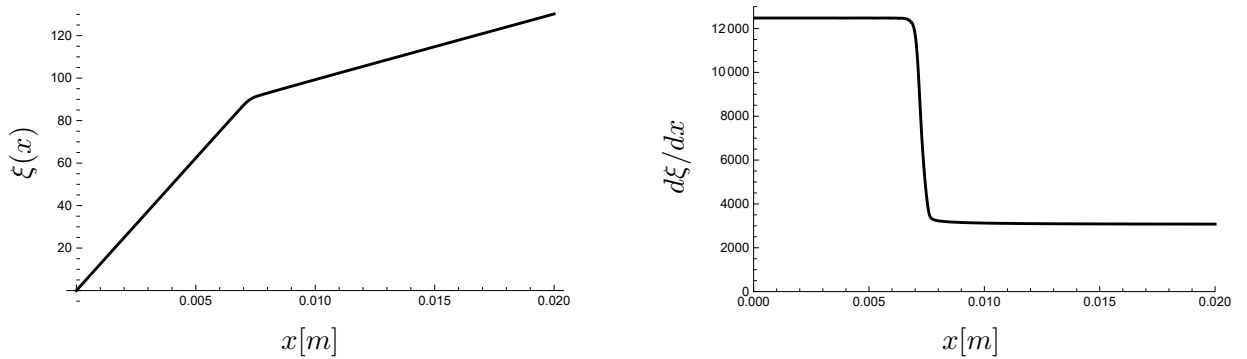

As a next step, we evaluated the canonical transformation from the physical coordinate $x$ to the canonical coordinate $\xi$ where  eq.(\ref{eq:cstrans}) is valid. Using the constant $\rho_u,s_L$ and the variable $(c_p/\lambda)(x)$ from the detailed chemistry free flame calculations, eq.(\ref{eq:xitrans}) is integrated numerically yielding a $\xi(x)$ shown in fig.(\ref{fig:xidxix}) for the case $\phi=1.0$.

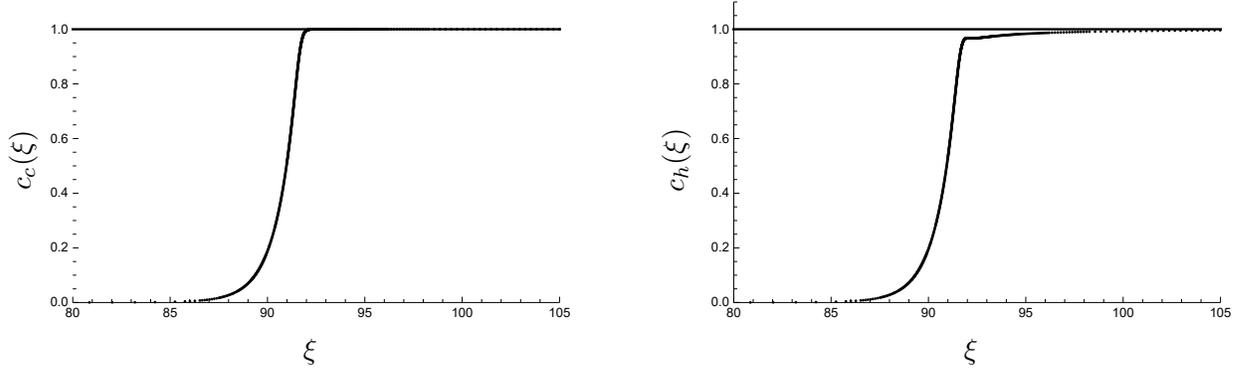
\begin{figure} [ht]
	\begin{minipage}[b]{.42\linewidth} % [b] => Ausrichtung an \caption
		\begin{tikzpicture}
			\node[] (Grafik) at (0,0) {\includegraphics[width=1\textwidth]{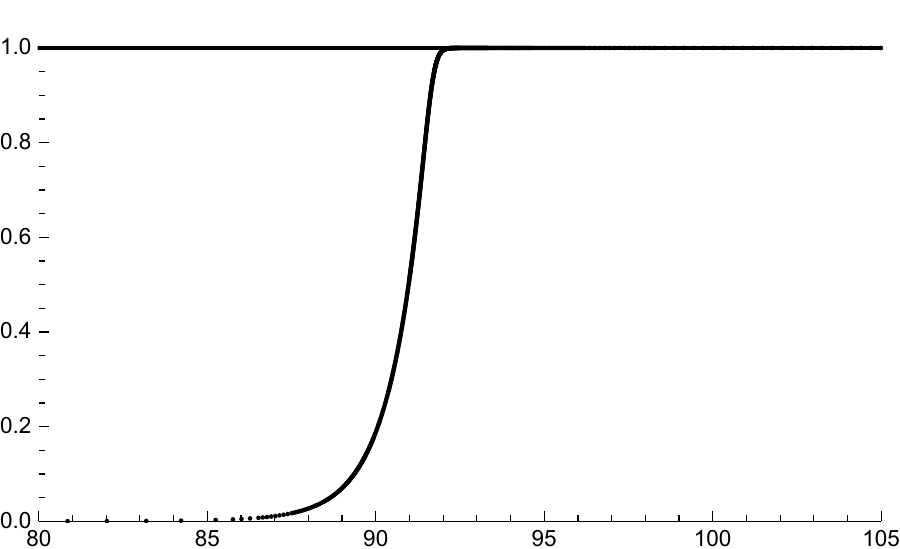}};
			\node[anchor=north,yshift=0pt,xshift=0] at (Grafik.south) {$\xi$};
			\node[rotate=90,anchor=south,xshift=0pt,yshift=-3] at (Grafik.west) {$c_c(\xi)$};
		\end{tikzpicture}
	\end{minipage}
	\hspace{.1\linewidth}% Abstand zwischen Bilder
	\begin{minipage}[b]{.42\linewidth} % [b] => Ausrichtung an \caption
		\begin{tikzpicture}
			\node[] (Grafik) at (0,0) { \includegraphics[width=\linewidth]{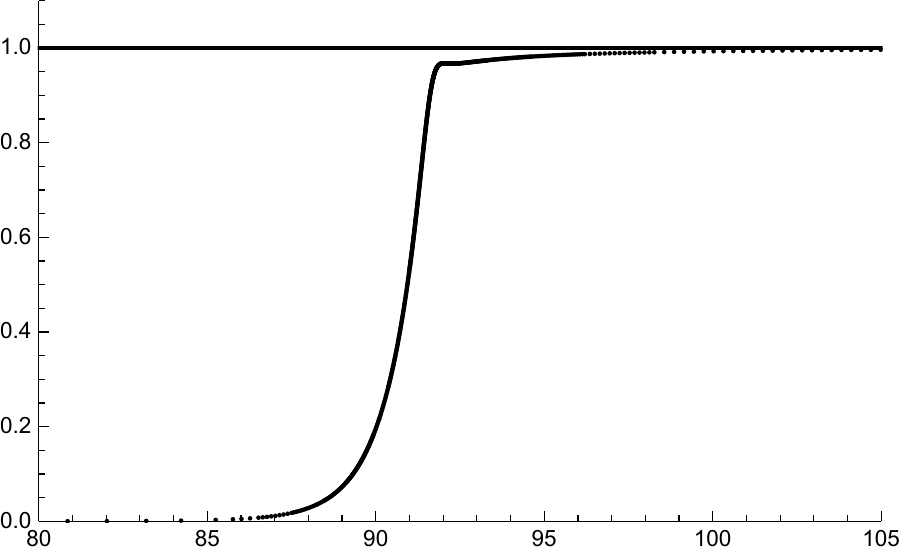}};
			\node[anchor=north,yshift=-0pt,xshift=0] at (Grafik.south) {$\xi$};
			\node[rotate=90,anchor=south,yshift=-0,xshift=0pt] at (Grafik.west) {$c_h(\xi)$};
		\end{tikzpicture} 
	\end{minipage}
	\caption{Normalized specific mole number of $CO_2+CO$ (left) and  $H_2O+H_2$ (right) in transformed coordinate $\xi$; free flame at $\phi=1$}
	\label{fig:tcnorphi1}
\end{figure}

One can see that the derivative of $\xi(x)$ drops by factor of $~4$ across the flame front mostly due to increase of heat conductivity as temperature increases. The normalized  $CO_2+CO$   and $H_2O+H_2$ specific mole number profiles in transformed coordinate $\xi$ are shown in fig.(\ref{fig:tcnorphi1}). Through the spatial coordinate transformation the thickness of the preheat zone  in $\xi$ space becomes more pronounced compared to the profiles in $x$ space.

Assuming that the $c$ profiles in $\xi$ coordinates fulfill eq.(\ref{eq:cstrans}), we can derive equivalent effective source term $\omega(c)$ by numerically evaluating first and second derivatives of $c(\xi$) and forming $\omega(\xi)=\partial c/\partial \xi - \partial^2 c / \partial \xi^2$. Parametric plots of $\omega(\xi)$ vs. $c(\xi)$ yields the effective $\omega(c)$

\begin{figure} [ht]
	\begin{minipage}[b]{.42\linewidth} % [b] => Ausrichtung an \caption
		\begin{tikzpicture}
			\node[] (Grafik) at (0,0) {\includegraphics[width=1\textwidth]{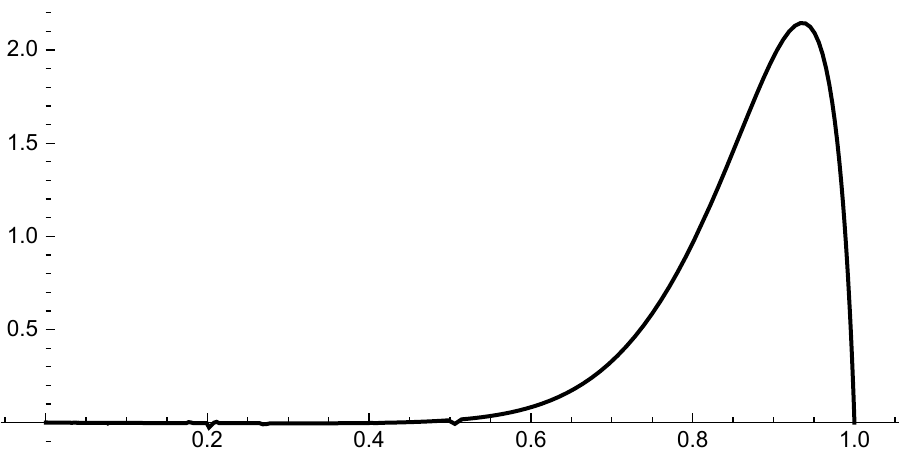}};
			\node[anchor=north,yshift=0pt,xshift=0] at (Grafik.south) {$c$};
			\node[rotate=90,anchor=south,xshift=0pt,yshift=-3] at (Grafik.west) {$\omega(c)$};
		\end{tikzpicture}
	\end{minipage}
	\hspace{.1\linewidth}% Abstand zwischen Bilder
	\begin{minipage}[b]{.42\linewidth} % [b] => Ausrichtung an \caption
		\begin{tikzpicture}
			\node[] (Grafik) at (0,0) { \includegraphics[width=\linewidth]{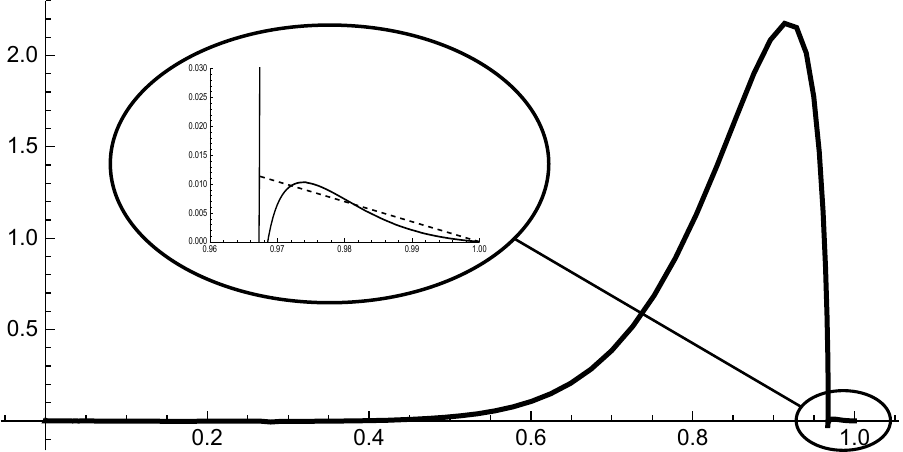}};
			\node[anchor=north,yshift=-0pt,xshift=0] at (Grafik.south) {$c$};
			\node[rotate=90,anchor=south,yshift=-0,xshift=0pt] at (Grafik.west) {$\omega(c)$};
		\end{tikzpicture} 
	\end{minipage}
	\caption{Numerically generated  $\omega(c)$ for $CO_2+CO$ (left) and  $H_2O+H_2$ (right); free flame at $\phi=1$}
	\label{fig:omnumtc}
\end{figure}

Such plots are shown in fig.(\ref{fig:omnumtc}) for $CO_2+CO$ and $H_2O+H_2$ progress variables. While $\omega(c)$ for $CO_2+CO$ looks indeed very similar to the single-step one from Fig.(\ref{fig:cromm8}), in case of $H_2O+H_2$ the one-step chemistry region appears to end at an $c=C \approx 0.968$. 

In the region between $c=C$ and $c=1$, $\omega(c)$ for $H_2O+H_2$ can be roughly approximated by a linear function  $\omega(c) \approx B (1-c)$. Such a linear source term\cite{ferziger1993simplified} also yields an analytical and invertible flame profile and pdf. The corresponding $c(\xi)$ solution of eq.(\ref{eq:cstrans}) is
\begin{equation}
	c(\xi)=1 + c_1 \cdot exp\left[\frac{1}{2}(1 - \sqrt{1 + 4 B}) \xi\right]
	\label{clinsource}
\end{equation}
discarding the term rising exponentially in $\xi$. $c_1$ is chosen for a $c(\xi)$ to be continuous at $c=C$ and $B$ can be selected to fit $\omega(c)$ in the region $C<c \leq 1$, see insert in fig(\ref{fig:omnumtc}). For the pdf in this region we need $dc/d\xi$ as function of $c$. It evaluates as $dc/d\xi=\frac{1}{2} \left(\sqrt{4B+1}-1\right) (1-c)$. When using a progress variable with a tail, the $c$ profile and pdf are to be defined separately in the range $0 \leq c \leq C$ and $C \leq c \leq 1$. This makes calculations slightly more complicated,  but the solutions are still analytical.

\section{Direct fit of $c(\xi)$ profiles}
A direct least squares fit $c_m(\xi-\xi_0)$ to the GRI mech 3.0 distributions of $CO_2+CO$ and $H_2O+H_2$ for $\phi=1$ yields $a \approx 1$ in both cases and $m=8.75$ and $m=9.65$, respectively. The shift $\xi_0$ is irrelevant to the profile shape since the position of the GRI mech 3.0 profiles in $x$ space is arbitrary. 

Note that the preheat zone will only be reproduced accurately (with $a=1$) with a consistent scaling factor $\rho_u s_L c_p / \lambda$. For example, the change of $\rho_u$ due to admixture of methan into air has to be considered properly. Fig.(\ref{fig:diffchfitphi10}) shows only the difference between the fitted $c_m(\xi)$ and the GRI mech 3.0 profiles for $\phi=1.0$ since an overlaid plot of fitted and GRI mech 3.0 profiles would not show any differences. The maximum difference is below $0.1 \%$ in both cases. 

Using the normalized temperature as a progress variable, the fit yielded $m=4.7, a=0.95$, indicating that for this progress variable a much smaller effective activation temperature is required. The slightly smaller value of stretch parameter $a$ is caused here by the interaction between heat release and the varying effective molar weight within the preheat zone. The difference between the fitted $c_m(\xi)$ and the GRI mech 3.0 is similar as in the $CO_2+CO$ and the $H_2O+H_2$ cases in the preheat / reaction regions up to $c=C=0.85$. The linear fit to $\omega(c)$ in the tail region $C \leq c \leq 1$  can also be applied here, but the difference between GRI mech 3.0 and fitted profile rises up to $1 \%$ in this region.

Similar levels reproduction of the GRI mech 3.0 profiles by $c_m(\xi)$ are achieved for lean flames down to $\phi=0.4$ and for rich flames up to $\phi=1.3$, see fig.(\ref{fig:diffchfitphi06}) for the case $\phi=0.6$.  Away from $\phi=1$, parameter $m$ rises slightly, indicating that slightly different single-step Arrhenius effective activation temperatures are necessary for flames at different $\phi$. 

\begin{figure} [ht]
	\begin{minipage}[b]{.42\linewidth} % [b] => Ausrichtung an \caption
		\begin{tikzpicture}
			\node[] (Grafik) at (0,0) {\includegraphics[width=1\textwidth]{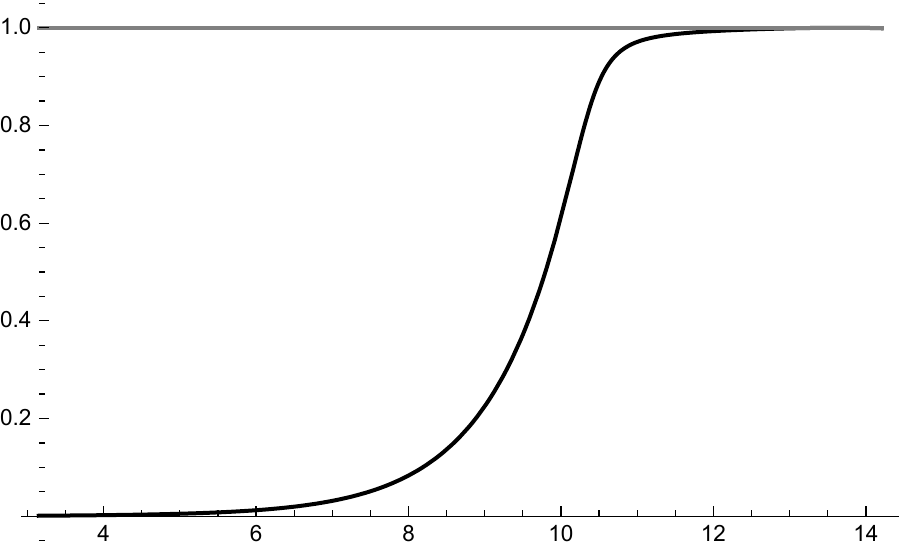}};
			\node[anchor=north,yshift=0pt,xshift=0] at (Grafik.south) {$\xi$};
			\node[rotate=90,anchor=south,xshift=0pt,yshift=-3] at (Grafik.west) {$c_c(\xi)$};
		\end{tikzpicture}
	\end{minipage}
	\hspace{.1\linewidth}% Abstand zwischen Bilder
	\begin{minipage}[b]{.42\linewidth} % [b] => Ausrichtung an \caption
		\begin{tikzpicture}
			\node[] (Grafik) at (0,0) { \includegraphics[width=\linewidth]{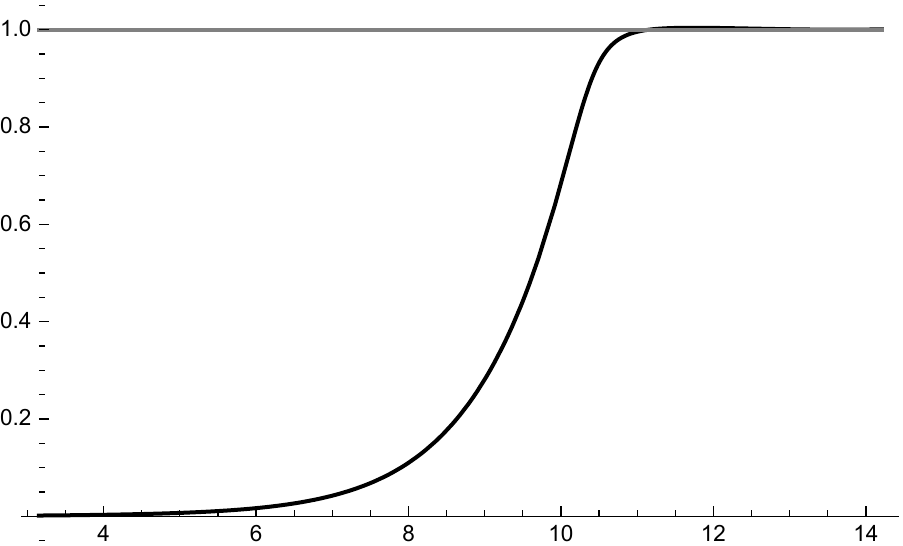}};
			\node[anchor=north,yshift=-0pt,xshift=0] at (Grafik.south) {$\xi$};
			\node[rotate=90,anchor=south,yshift=-0,xshift=0pt] at (Grafik.west) {$c_t(\xi)$};
		\end{tikzpicture} 
	\end{minipage}
	\caption{Normalized specific mole number of $CO_2+CO$ (left) and  temperature (right) in transformed coordinate $\xi$; free flame at $\phi=2$}
	\label{fig:tcnorphirich}
\end{figure}

For very rich flames (here $1.3 \leq \phi \leq 2.2$), the $CO_2+CO$ specific mole number profiles develop a small long tail towards $c=1$, while the tail tends to disappear from the normalized temperature profiles. Fig.(\ref{fig:tcnorphirich}) shows transformed normalized $CO_2+CO$ and temperature profiles for $\phi=2$. Since for Lewis numbers equal to one the transport equations for species-derived progress variables and for normalized temperature are identical, a tabulation of flamelet quantities vs. progress variable might use a $CO_2+CO$ progress variable for $\phi<1.3$ and switch to a table based on normalized temperature for richer flames.

\begin{figure} [ht]
	\begin{minipage}[b]{.42\linewidth} % [b] => Ausrichtung an \caption
		\begin{tikzpicture}
			\node[] (Grafik) at (0,0) {\includegraphics[width=1\textwidth]{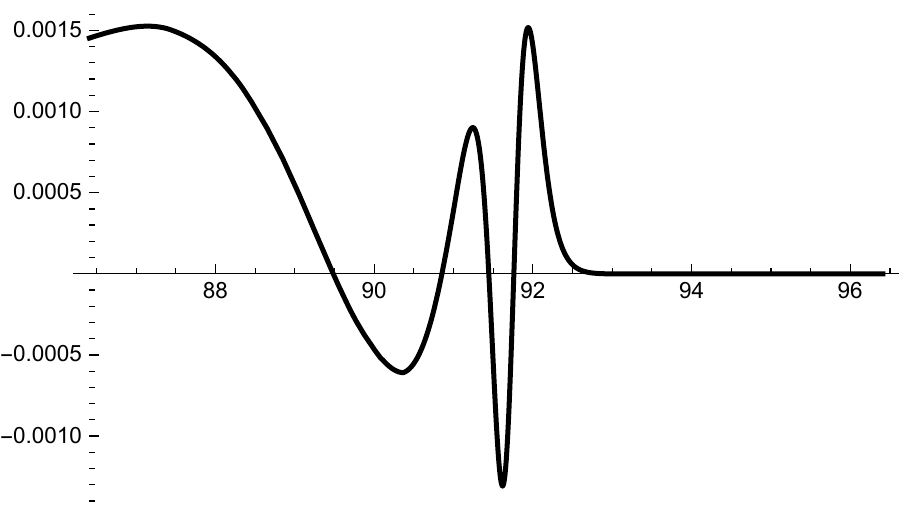}};
			\node[anchor=north,yshift=0pt,xshift=0] at (Grafik.south) {$\xi$};
			\node[rotate=90,anchor=south,xshift=0pt,yshift=-3] at (Grafik.west) {$c(\xi)-c_m(\xi)$};
		\end{tikzpicture}
	\end{minipage}
	\hspace{.1\linewidth}% Abstand zwischen Bilder
	\begin{minipage}[b]{.42\linewidth} % [b] => Ausrichtung an \caption
		\begin{tikzpicture}
			\node[] (Grafik) at (0,0) { \includegraphics[width=\linewidth]{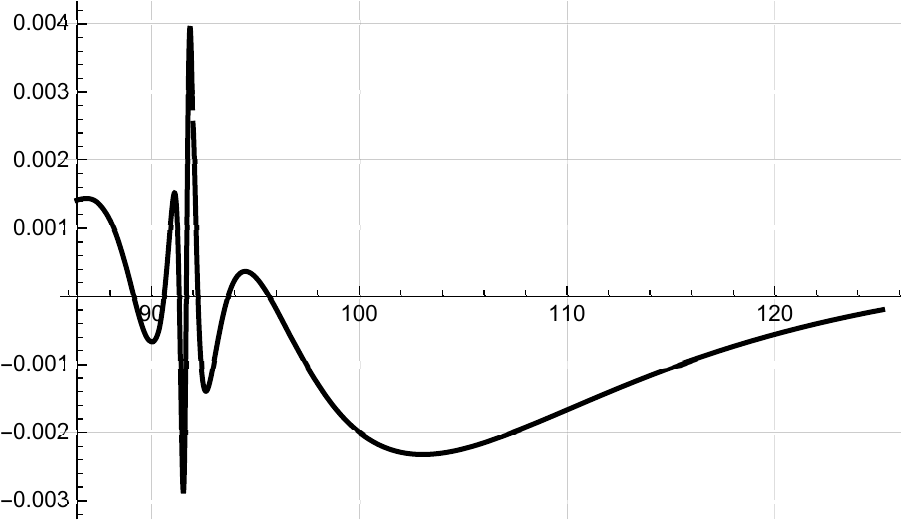}};
			\node[anchor=north,yshift=-0pt,xshift=0] at (Grafik.south) {$\xi$};
			\node[rotate=90,anchor=south,yshift=-0,xshift=0pt] at (Grafik.west) {$c(\xi)-c_m(\xi)$};
		\end{tikzpicture} 
	\end{minipage}
	\caption{Difference between fitted and GRI mech 3.0 profile  for $CO_2+CO$ (left) and  $H_2O+H_2$ (right); free flame at $\phi=1$}
	\label{fig:diffchfitphi10}
\end{figure}
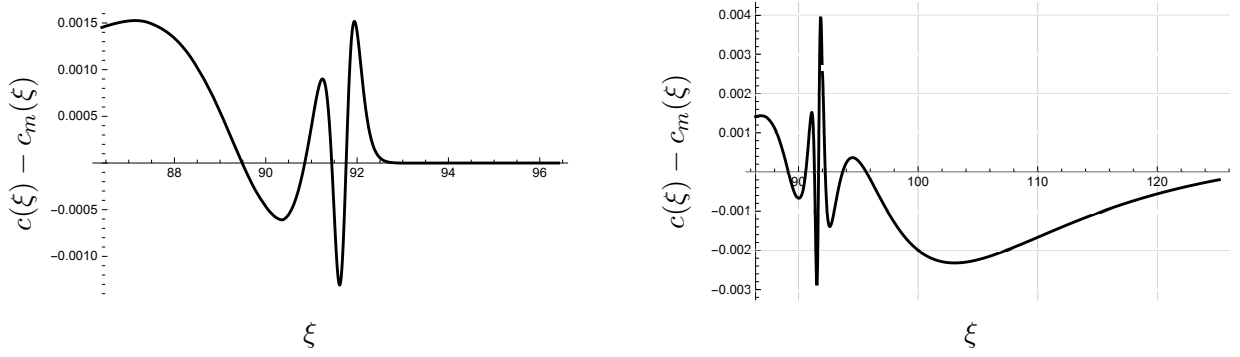
\begin{figure} [ht]
	\begin{minipage}[b]{.42\linewidth} % [b] => Ausrichtung an \caption
		\begin{tikzpicture}
			\node[] (Grafik) at (0,0) {\includegraphics[width=1\textwidth]{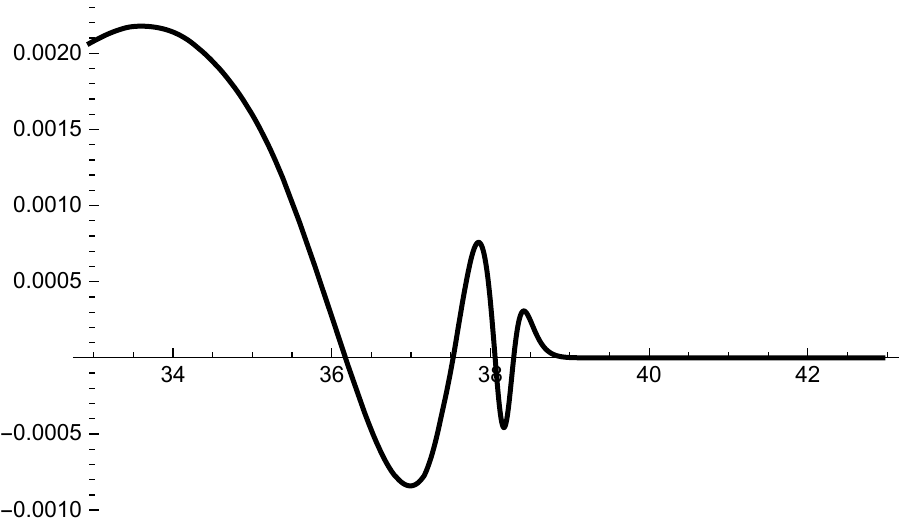}};
			\node[anchor=north,yshift=0pt,xshift=0] at (Grafik.south) {$\xi$};
			\node[rotate=90,anchor=south,xshift=0pt,yshift=-3] at (Grafik.west) {$c(\xi)-c_m(\xi)$};
		\end{tikzpicture}
	\end{minipage}
	\hspace{.1\linewidth}% Abstand zwischen Bilder
	\begin{minipage}[b]{.42\linewidth} % [b] => Ausrichtung an \caption
		\begin{tikzpicture}
			\node[] (Grafik) at (0,0) { \includegraphics[width=\linewidth]{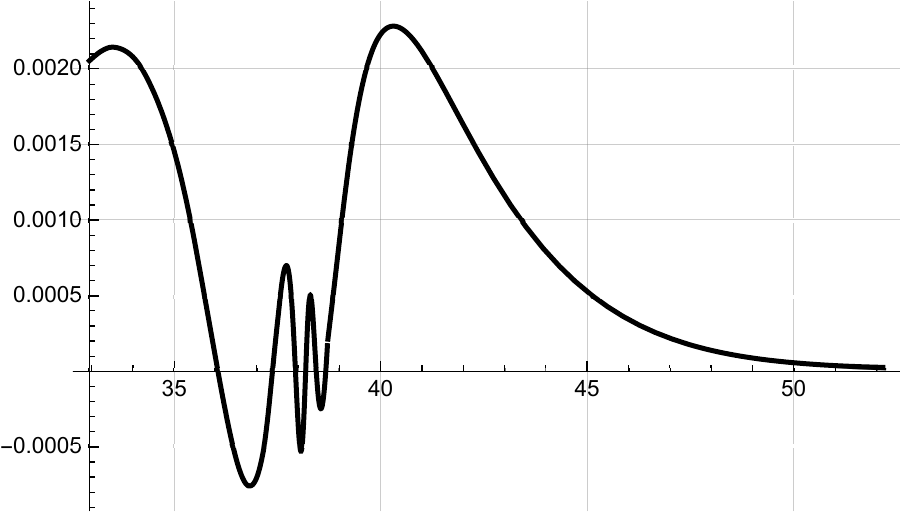}};
			\node[anchor=north,yshift=-0pt,xshift=0] at (Grafik.south) {$\xi$};
			\node[rotate=90,anchor=south,yshift=-0,xshift=0pt] at (Grafik.west) {$c(\xi)-c_m(\xi)$};
		\end{tikzpicture} 
	\end{minipage}
	\caption{Difference between fitted and GRI mech 3.0 profile  for $CO_2+CO$ (left) and  $H_2O+H_2$ (right); free flame at $\phi=0.6$}
	\label{fig:diffchfitphi06}
\end{figure}

\section{Alternative determination of parameter m}
In this section we focus on an alternative derivation of parameter $m$ using $T_u$,$T_b(\phi)$ and $s_L(\phi)$ from the GRI mech 3.0 calculations only. We only present results for the $CO_2+CO$ progress variable, the application of this method to the $H_2O+H_2$ and the normalized temperature progress variables yields similar results.

In recent contributions \cite{pfitzner2020pdf}$^,$\cite{pfitzner2021pdf} we have derived analytical results for the laminar flame Eigenvalue $\Lambda$ in the case of single-step Arrhenius chemistry. We have also provided relations to evaluate the profile parameter $m$ from the Arrhenius parameters $\alpha, \beta, \beta_1$.  
\begin{equation}
	m=\frac{4}{5}(\alpha + \beta) - 1
	\label{eq:m0}
\end{equation}
\begin{equation}
	\Lambda=\frac{\beta^2}{2}(\frac{1 - \alpha}{100} + 1) + 2\alpha\beta - \frac{22}{15}\beta
	\label{eq:Lam0}
\end{equation}
Since results for $\beta_1=0$ and $\beta_1=1$ were very similar, we only present results for $\beta_1=0$.  
\begin{figure} [ht]
	\begin{minipage}[b]{.42\linewidth} % [b] => Ausrichtung an \caption
		\begin{tikzpicture}
			\node[] (Grafik) at (0,0) {\includegraphics[width=1\textwidth]{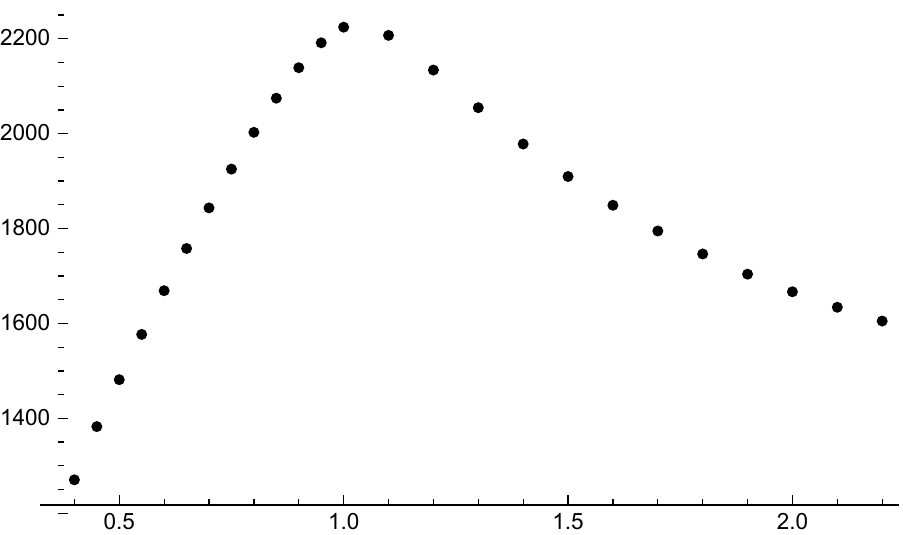}};
			\node[anchor=north,yshift=0pt,xshift=0] at (Grafik.south) {$\phi$};
			\node[rotate=90,anchor=south,xshift=0pt,yshift=-3] at (Grafik.west) {$T_b$};
		\end{tikzpicture}
	\end{minipage}
	\caption{$T_b$ from GRI mech 3.0 free flame calculations}
	\label{fig:tmaxsl}
\end{figure}

Using $\alpha=(T_b-T_u)/T_b$ and $\beta=\alpha T_a/T_b$, we can first estimate an effective activation temperature of $T_a \approx 29200 K$ for $\phi=1$ using eq.(\ref{eq:m0}) and $m=8.75$ from the fit to the $\phi=1$ free flame profile. $T_u=\,300\,K$ and $T_b$ is taken from the GRI mech 3.0 profiles, see fig.(\ref{fig:tmaxsl}).  Note that $T_b$ could also be calculated from equilibrium thermodynamics, since the fully burnt state of free flames should approach the thermodynamic equilibrium state.

For Arrhenius single step chemistry, the laminar flame speed is\cite{poi05}
\begin{equation}
	s_L \propto \sqrt{T_b exp(-\frac{\beta}{\alpha}) \Lambda }
\end{equation}
up to constants independent of $\phi$. 
For constant activation temperature $T_a$, we can predict the ratio $s_L(\phi)/s_L(\phi=1)$. The results are shown in fig.(\ref{fig:slmfit}, left). We find that these predictions of $s_L(\phi)/s_L(\phi=1)$ agree qualitatively with the GRI mech 3.0 results but the precise value of $s_L$ is very sensitive on the activation temperature $T_a$ or equivalently, $m$. 

We therefore suggest to reverse the procedure, calculating an effective $T_a$ or equivalently $m$ for $\phi \neq 1$ from the GRI mech 3.0 $s_L(\phi)/s_L(1)$ and $T_b(\phi)$. The resulting $m$ distribution is shown in fig.(\ref{fig:slmfit}, right) together with the $m$ distribution evaluated for constant $T_a$. The slight change of $m$ between the two distributions generate profiles which are hardly distinguishable. Obviously, using the adapted $m's$ from fig. (\ref{fig:slmfit}, right) would bring the gray symbols in fig. (\ref{fig:slmfit}, left) exactly on top of the black ones. 

The $c(\xi)$ profiles evaluated with $m$ determined in this way also agree still very well with the detailed chemistry profiles, see fig.(\ref{fig:cmslfitmphi06}). The difference is still below $0.4 \%$ and invisible in the plot on the left side. For use in a CFD code, $s_L(\phi)$ and $m(\phi)$ calculated from $s_L(\phi),T_b(\phi)$  can easily be represented by polynomials in $\phi$.
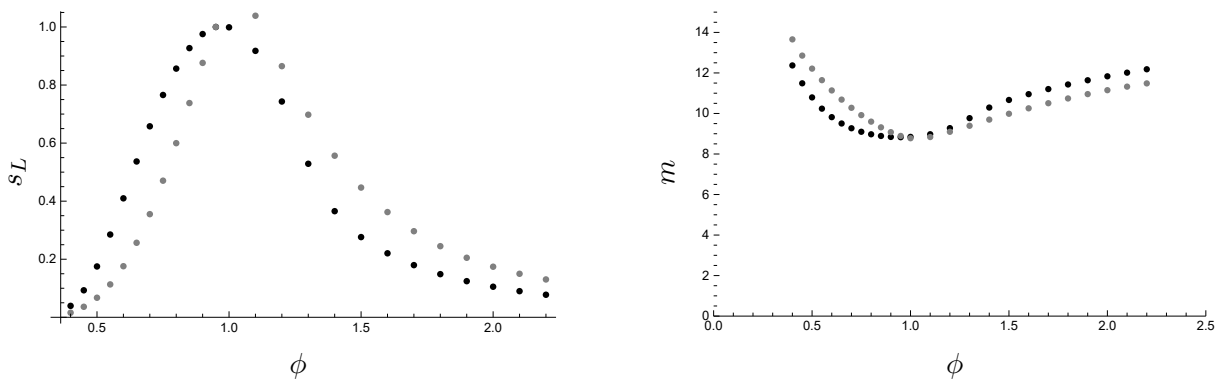
\begin{figure} [ht]
	\begin{minipage}[b]{.42\linewidth} % [b] => Ausrichtung an \caption
		\begin{tikzpicture}
			\node[] (Grafik) at (0,0) {\includegraphics[width=1\textwidth]{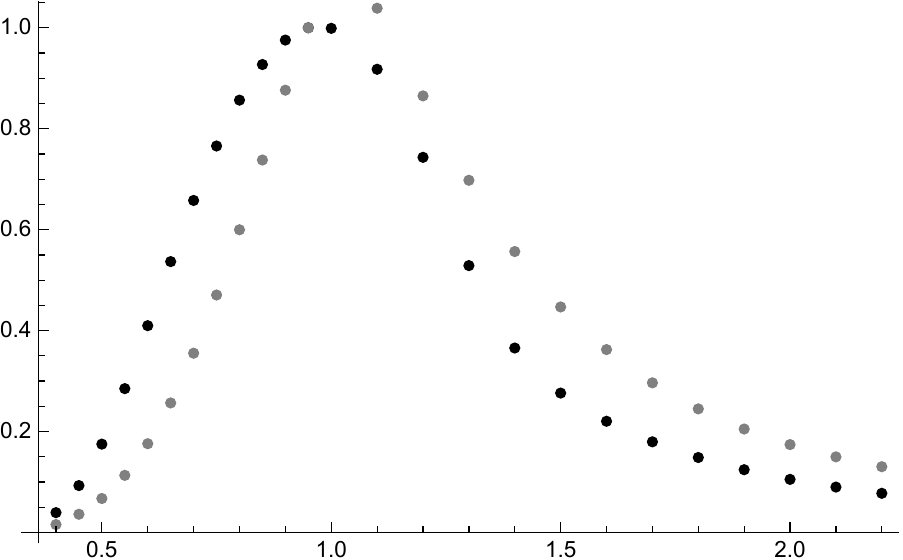}};
			\node[anchor=north,yshift=0pt,xshift=0] at (Grafik.south) {$\phi$};
			\node[rotate=90,anchor=south,xshift=0pt,yshift=-3] at (Grafik.west) {$s_L$};
		\end{tikzpicture}
	\end{minipage}
	\hspace{.1\linewidth}% Abstand zwischen Bilder
	\begin{minipage}[b]{.42\linewidth} % [b] => Ausrichtung an \caption
		\begin{tikzpicture}
			\node[] (Grafik) at (0,0) { \includegraphics[width=\linewidth]{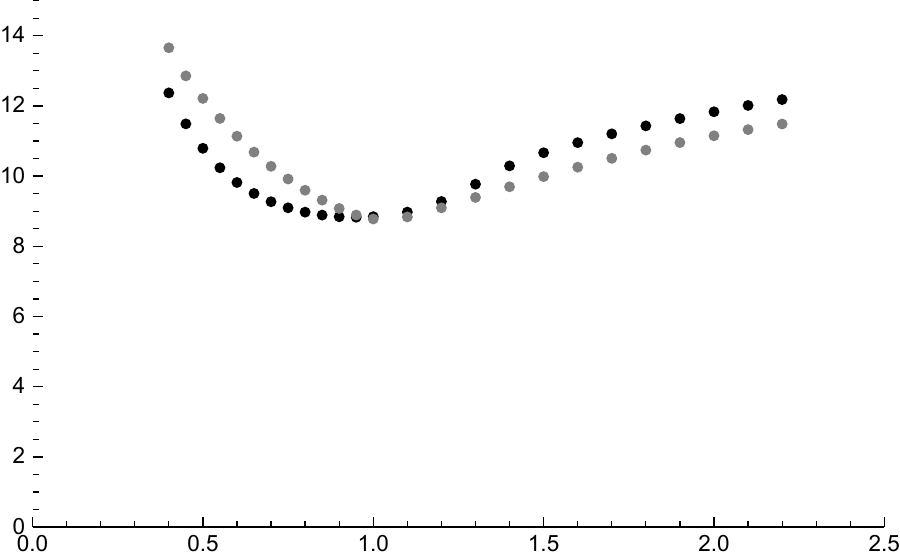}};
			\node[anchor=north,yshift=-0pt,xshift=0] at (Grafik.south) {$\phi$};
			\node[rotate=90,anchor=south,yshift=-0,xshift=0pt] at (Grafik.west) {$m$};
		\end{tikzpicture} 
	\end{minipage}
	\caption{Left: $s_L/s_{L,\phi=1}$ from GRI mech 3.0 calculations (black) and calculated using constant $T_a$ (gray); right: $m$ for constant $T_a$ (gray) and predicted using GRI mech 3.0 $s_L/s_{L,\phi=1}$ (black)}
	\label{fig:slmfit}
\end{figure}
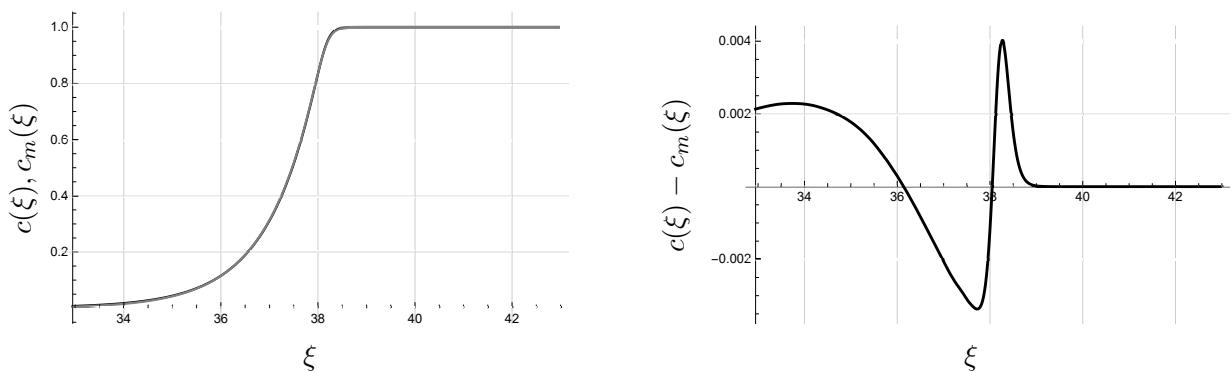
\begin{figure} [ht]
	\begin{minipage}[b]{.42\linewidth} % [b] => Ausrichtung an \caption
		\begin{tikzpicture}
			\node[] (Grafik) at (0,0) {\includegraphics[width=1\textwidth]{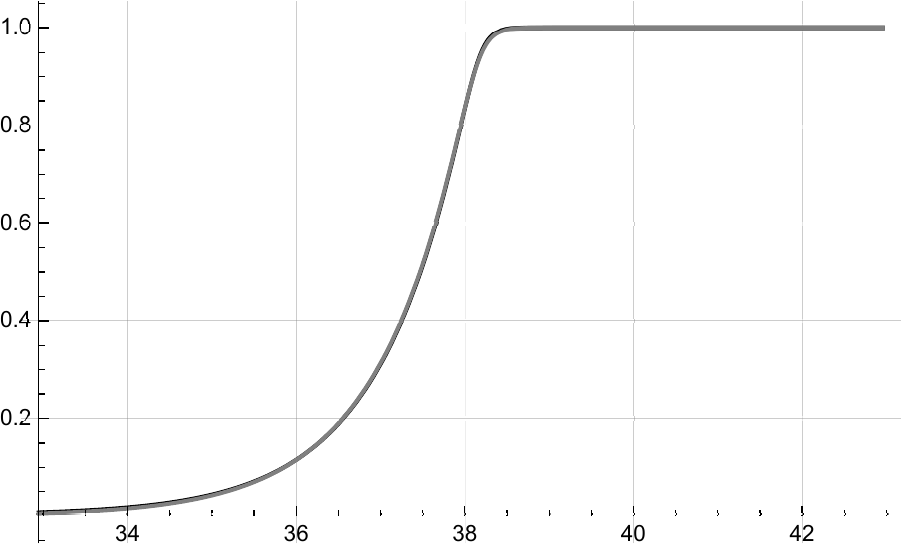}};
			\node[anchor=north,yshift=0pt,xshift=0] at (Grafik.south) {$\xi$};
			\node[rotate=90,anchor=south,xshift=0pt,yshift=-3] at (Grafik.west) {$c(\xi),c_m(\xi)$};
		\end{tikzpicture}
	\end{minipage}
	\hspace{.1\linewidth}% Abstand zwischen Bilder
	\begin{minipage}[b]{.42\linewidth} % [b] => Ausrichtung an \caption
		\begin{tikzpicture}
			\node[] (Grafik) at (0,0) { \includegraphics[width=\linewidth]{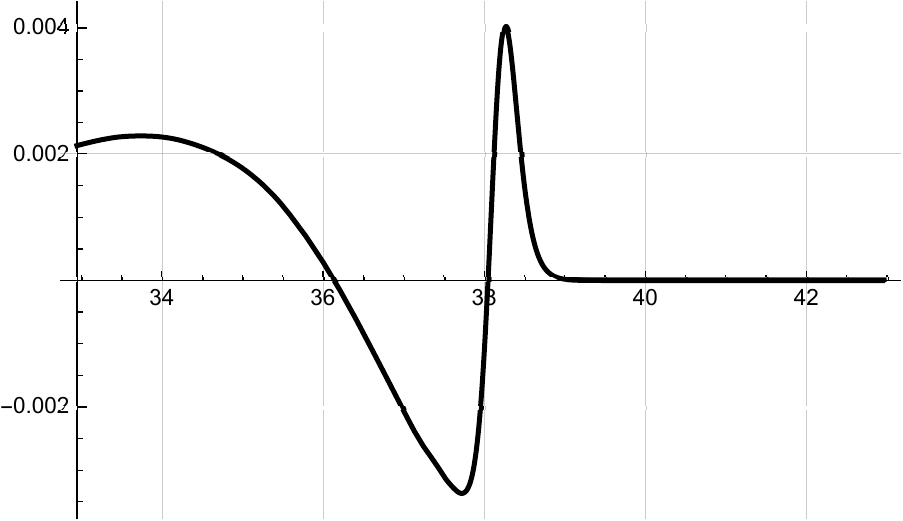}};
			\node[anchor=north,yshift=-0pt,xshift=0] at (Grafik.south) {$\xi$};
			\node[rotate=90,anchor=south,yshift=-0,xshift=0pt] at (Grafik.west) {$c(\xi)-c_m(\xi)$};
		\end{tikzpicture} 
	\end{minipage}
	\caption{Analytical (gray) and GRI mech 3.0 (black) $CO_2+CO$ profile (left) and  difference (right); free flame at $\phi=0.6$, $m$ determined from $s_L,T_b$}
	\label{fig:cmslfitmphi06}
\end{figure}

\section{Effect of strain}
\begin{figure} [ht]
 	\begin{minipage}[b]{.42\linewidth} % [b] => Ausrichtung an \caption
 		\begin{tikzpicture}
 			\node[] (Grafik) at (0,0) {\includegraphics[width=1\textwidth]{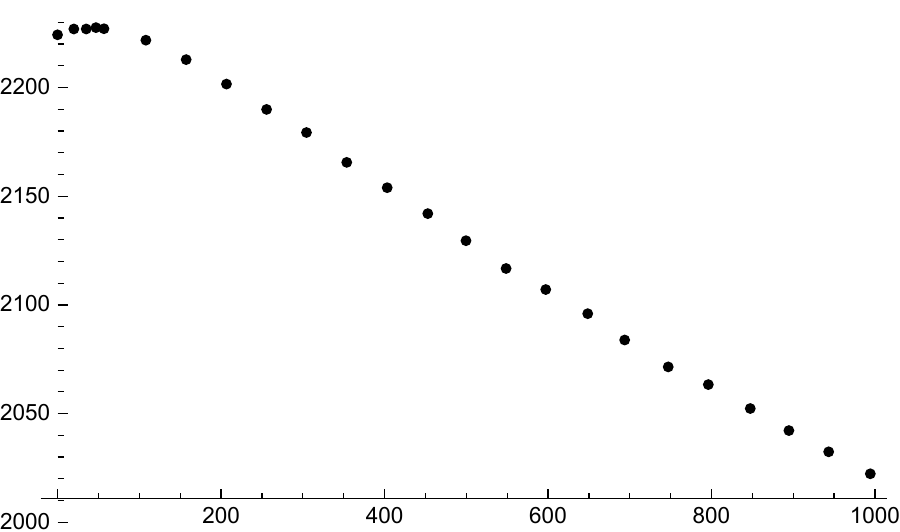}};
 			\node[anchor=north,yshift=0pt,xshift=0] at (Grafik.south) {$K$};
 			\node[rotate=90,anchor=south,xshift=0pt,yshift=-3] at (Grafik.west) {$T_b$};
 		\end{tikzpicture}
 	\end{minipage}
 	\hspace{.1\linewidth}% Abstand zwischen Bilder
 	\begin{minipage}[b]{.42\linewidth} % [b] => Ausrichtung an \caption
 		\begin{tikzpicture}
 			\node[] (Grafik) at (0,0) { \includegraphics[width=\linewidth]{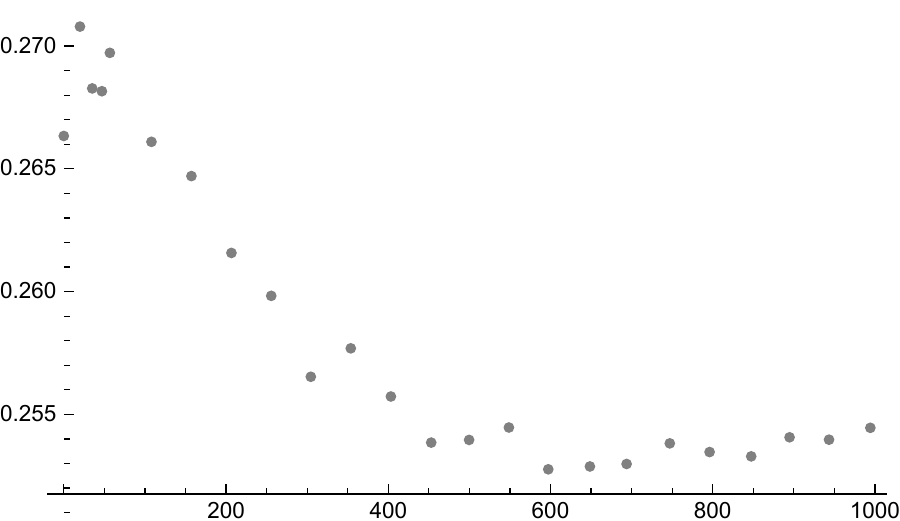}};
 			\node[anchor=north,yshift=-0pt,xshift=0] at (Grafik.south) {$K$};
 			\node[rotate=90,anchor=south,yshift=-0,xshift=0pt] at (Grafik.west) {$s_L$};
 		\end{tikzpicture} 
 	\end{minipage}
 	\caption{GRI mech 3.0 burnt temperature (left) and laminar flame speed of strained flames (right)}
 	\label{fig:tmaxslstrain}
 \end{figure}
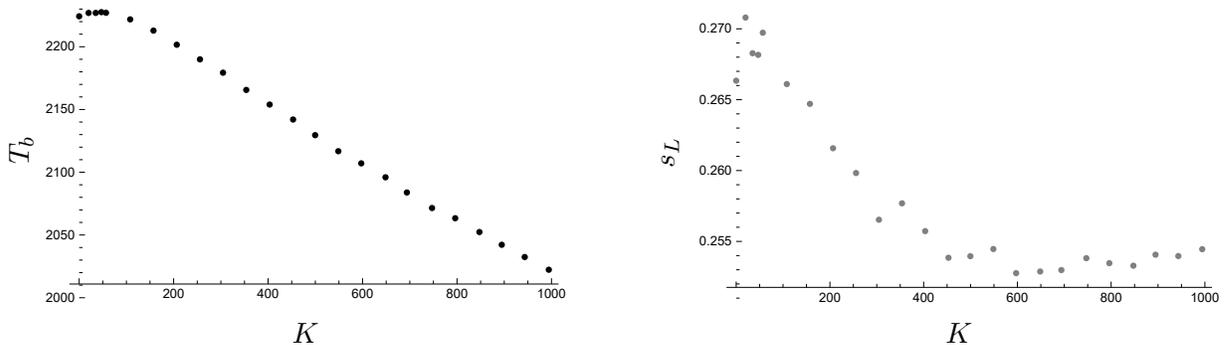

To investigate whether the effect of strain on $c$ profiles can also be represented by $c_m(\xi)$, we analysed a series of strained GRI mech 3.0 laminar premixed flames strained in a counterflow unburnt-unburnt configuration at $\phi=1$ and $\phi=0.6$. Results for both $\phi's$ were qualitatively similar, therefore we show only results for $\phi=1$. Fig.(\ref{fig:tmaxslstrain}) displays $T_b$ and $s_L$ as function of the strain rate $K$.

\begin{figure} [ht]
	\begin{minipage}[b]{.4\linewidth} % [b] => Ausrichtung an \caption
		\begin{tikzpicture}
			\node[] (Grafik) at (0,0) {\includegraphics[width=1\textwidth]{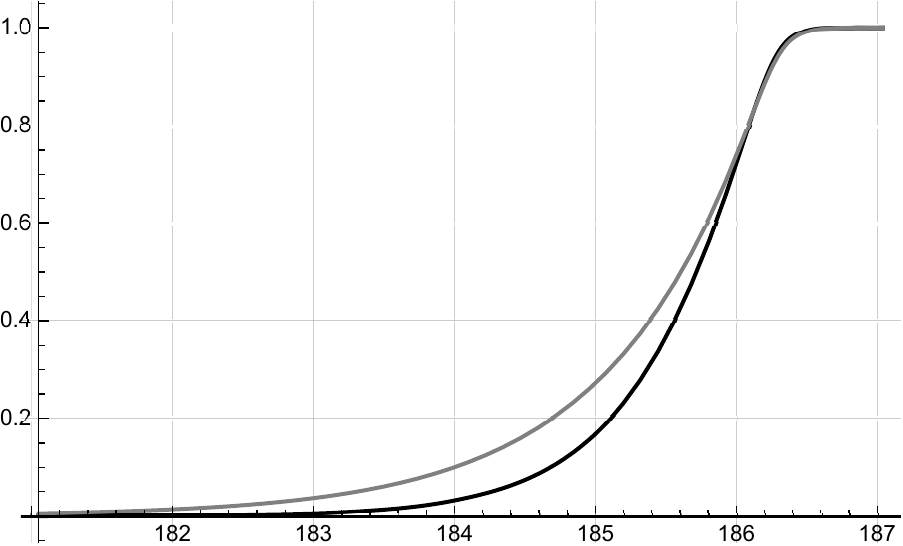}};
			\node[anchor=north,yshift=0pt,xshift=0] at (Grafik.south) {$\xi$};
			\node[rotate=90,anchor=south,xshift=0pt,yshift=-3] at (Grafik.west) {$c(\xi),c_m(\xi)$};
		\end{tikzpicture}
	\end{minipage}
	\hspace{.1\linewidth}% Abstand zwischen Bilder
	\begin{minipage}[b]{.4\linewidth} % [b] => Ausrichtung an \caption
		\begin{tikzpicture}
			\node[] (Grafik) at (0,0) { \includegraphics[width=\linewidth]{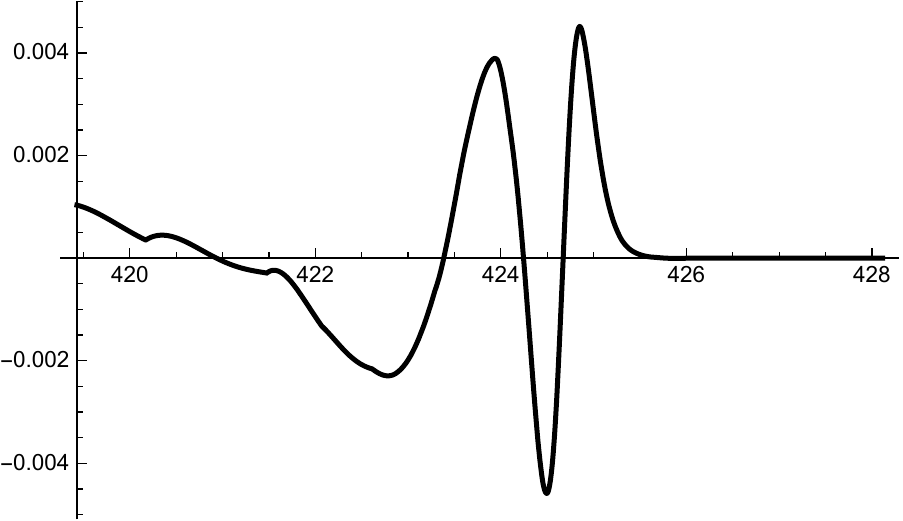}};
			\node[anchor=north,yshift=-0pt,xshift=0] at (Grafik.south) {$\xi$};
			\node[rotate=90,anchor=south,yshift=-0,xshift=0pt] at (Grafik.west) {$c(\xi)-c_m(\xi)$};
		\end{tikzpicture} 
	\end{minipage}
	\caption{Left: $c_m(\xi)$ with $a=1$ (gray) and GRI mech 3.0 strained $CO_2+CO$ profile (black); right: difference between scaled GRI mech 3.0 and fitted profiles, $a=1.5$; $\phi=0.6$, $K=688$}
	\label{fig:ccstrain}
\end{figure}
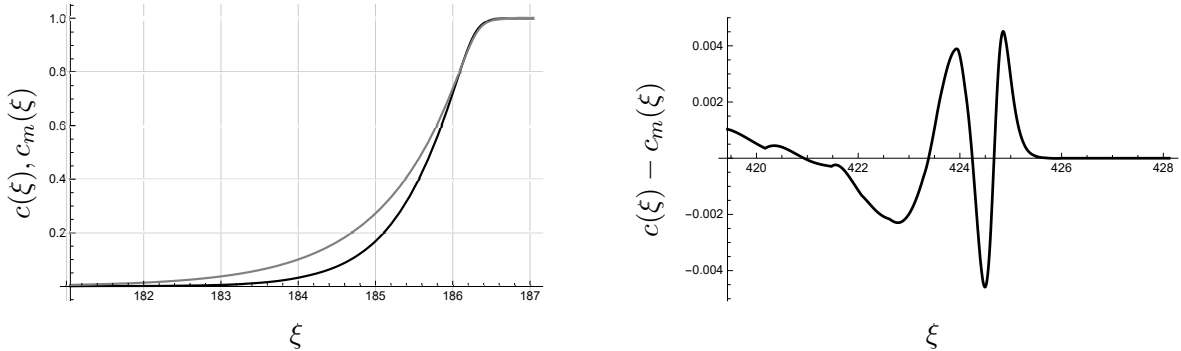

We find that for these strained flames, the thermal flame thickness (evaluated as inverse of the maximum derivative of the transformed detailed chemistry $c$ profiles)  is practically constant, independent of strain level $K$. This is consistent with laminar flame theory \cite{lipatnikov2012fundamentals} which holds that the reaction zone will be almost unaffected by levels of strain away from the blow-off level. However, the preheat zone will be compressed compared to a free flame situation. Fig.(\ref{fig:ccstrain}, left) shows the GRI mech 3.0 profile with an analytic profile of the same thermal flame thickness and without additional compression (i.e. $a=1$). It is evident that the preheat zone could not be reproduced by such a $c_m(\xi)$.

The strained profiles however can be reproduced well by $c_m(\xi)$ when fitting $m$ and $a$ simultaneously. Fig.(\ref{fig:ccstrain}, right) only shows the difference between the GRI mech 3.0 transformed profile and $c_m(\xi)$  for ($\phi=0.6, K=688$). A stretch factor of $a=1.50$ is required in this case to compress the preheat zone. The difference between GRI mech 3.0 and fitted profile is again within $0.4 \%$, as in the case of non-strained profiles. 

The stretch factor $a$ required for preheat zone compression can be correlated as $a=(1+5.5 \cdot 10^{-5} \cdot Ka) $ with the Karlowitz number $Ka=K \cdot \delta_{th}^0 / s_L^0$, with $\delta_{th}^0$ and $s_L^0$ evaluated from the unstrained flames. To retain the thermal thickness of the profiles for strained profiles with $a>1$, parameter $m$ has to decrease with rising $K$. The fitted $m$ values can be approximated as $m=m_{K=0}-\frac{3.9 \cdot K}{1000}$ for both $\phi=1.0$ and $\phi=0.6$.

\section{Premixed laminar flame pdf with detailed chemistry and non-constant $c_p/\lambda$}
The laminar flame pdf for constant $c_p/\lambda$ is given by eq.(\ref{eq:pdfdef}). For non-constant $c_p/\lambda$, the spatial region covered by a certain interval $d\xi$ is not constant, so spatial means in $\xi$ space are not proportional to spatial means in $x$ space.

A spatial mean of quantity $z$ in $x$ space for a filter of size $\Delta_x$ is defined as
\begin{equation}
	\overline{z(x)}=\frac{1}{\Delta_x}\int_{x}^{x+\Delta_x} z(x)dx
	\label{eq:zmean}
\end{equation}
Since $d\xi=\rho_u s_L c_p / \lambda dx$ we can define a scaling function $R(x)$ through
\begin{equation}
	\rho_u s_L (c_p / \lambda)=r_u/R(x)
	\label{eq:Rdef}
\end{equation}
where $r_u=\rho_u s_L (c_p / \lambda)_u$ is the stretch factor in the unburnt region. $R(x)$ is equal to one in this region and rises across the flame front.

\begin{figure} [ht]
\begin{minipage}[b]{.42\linewidth} % [b] => Ausrichtung an \caption
	\begin{tikzpicture}
		\node[] (Grafik) at (0,0) {\includegraphics[width=1\textwidth]{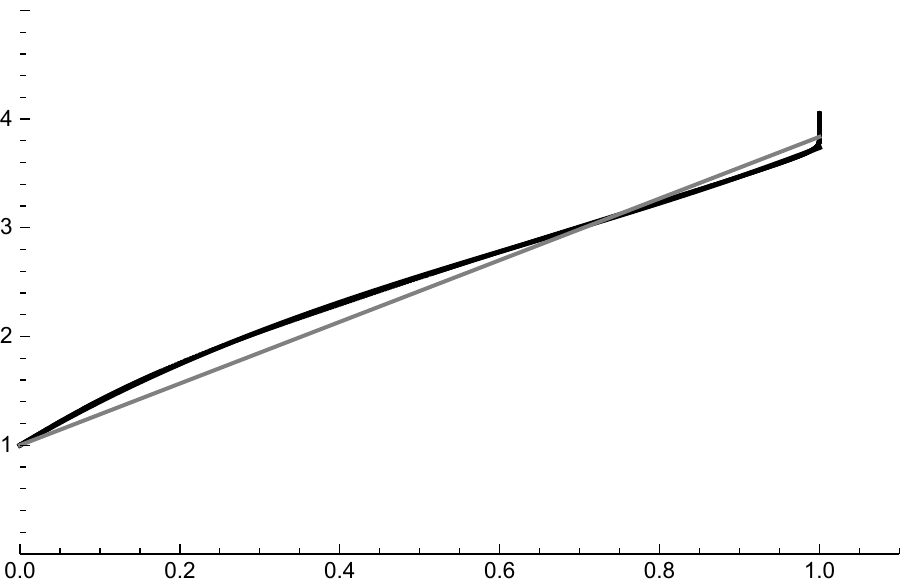}};
		\node[anchor=north,yshift=0pt,xshift=0] at (Grafik.south) {$c$};
		\node[rotate=90,anchor=south,xshift=0pt,yshift=-3] at (Grafik.west) {$R(c)$};
	\end{tikzpicture}
\end{minipage}
	\caption{Scaling factor as function of $c$ (black); also shown is linear approximation (gray)}
	\label{fig:Rxphi1}
\end{figure}

Since $c(x)$ is monotonous in $x$, $R(x)$ can be transformed into a $R(c)$, which is shown in fig.(\ref{fig:Rxphi1}) for
the $CO_2+CO$ progress variable. If the last (almost vertical) portion of $R(c)$ is discarded (it will not contribute to the mean value of $\omega$, which is zero at $c=1$), $R(c)$ can very accurately be represented by a $3^{rd}$ order polynomial in $c$, the constant term being equal to one. The simple linear approximation $R(c)=1+c \cdot \frac{7}{10} \cdot \frac{r_u}{r_b}$, also shown in fig.(\ref{fig:Rxphi1}), yields only slightly less accurate results.

The integral in eq.(\ref{eq:zmean}) can be evaluated as
\begin{equation}
	\frac{1}{\Delta_x}\int_{x}^{x+\Delta_x} z(x)dx=\frac{r_u}{\Delta_\xi}\int_{\xi^-}^{\xi^+} \frac{z(\xi)}{d\xi/dx}d\xi=\frac{1}{\Delta_\xi}\int_{c^-}^{c^+} \frac{z(c) R(c)}{dc/d\xi}dc
	\label{eq:Rdef}
\end{equation}
where $\Delta_\xi=\Delta_x \cdot r_u$ is to be evaluated using the stretch factor in the unburnt region. The integration limits $\xi^-,\xi^+,c^-,c^+$ are determined from $x,\Delta_x$ through $\xi^-=\xi(x)\,,\xi^+=\xi(x+\Delta_x)\,,c^-=c(\xi^-)$ and $c^+=c(\xi^+)$. Note that with nonlinear stretching,  $\xi^+ \neq \xi^-+\Delta_\xi$ with the above definition of $\Delta_\xi$.

For non-constant stretch factor the laminar flame pdf becomes therefore
\begin{equation}
		p(c) = \frac{1}{\Delta_\xi}\frac{R(c)}{dc/d\xi}H(c-c^-)H(c^+-c)
	\label{eq:pdfdefR}
\end{equation}
Inserting $dc/d\xi$ as function of $c$ from eq.(\ref{eq:dcdxi}), we see that for polynomial $R(c)$, all integrals required to evaluate $\overline{p(c)}=1$, $\overline{c}$ and $\overline{\omega_m(c)}$ can be evaluated analytically using
\begin{equation}
	I_0(c,m)=\int\frac{dc}{c(1-(c/C)^m)}=\log (c)-\frac{\log \left(1-\left(\frac{c}{C}\right)^n\right)}{n}
	\label{eq:i1c}
\end{equation}
\begin{equation}
	I_k(c,m) = \int\frac{c^k dc}{c(1-(c/C)^m)}=\frac{c^k \, _2F_1\left(1,\frac{k}{m};\frac{k}{m}+1;\left(\frac{c}{C}\right)^m\right)}{k}
	\label{eq:ikc}
\end{equation}
where $_2F_1(a,b,c;x)$ is a hypergeometric function and $k \neq 0$ in eq.(\ref{eq:ikc}). Note that $I_k(c,m)=\frac{I_1(c^k,m/k)}{k}$, so for numerical evaluation only $I_1(c,m)$ is needed; a robust method\cite{pfitzner2020pdf} for numerical evaluation of this function was provided earlier. 

Note also that for polynomical $R(c)$, $\frac{\omega_m(c)}{dc/d\xi}$ is a polynomial, see eq.(\ref{eq:intomegaCa}), even for detailed chemistry $c$ profiles.
In the case of the $CO_2+CO$ progress variable and ignoring the effect of strain for the evaluation of the mean source term (since strain will only
affect the preheat region where $\omega$ is small), we have $a=C=1$. Writing the polynomial expansion of $R(c)$ as
\begin{equation}
	R(c) = 1+\sum_{k=1}^{r} R_k c^k
	\label{eq:Rcpoly}
\end{equation}
we obtain
\begin{equation}
	1 = \int_0^1 p(c)dc=\frac{1}{N}\int_{c^-}^{c^+} \frac{R(c)}{c(1-c^m)}=\frac{1}{N}\left[\log (c)-\frac{\log \left(1-c^n\right)}{n}+\sum_{k=1}^{r} R_k I_{k}(c,m) \right]_{c^-}^{c^+}
	\label{eq:intpcstretch}
\end{equation} 
\begin{equation}
	\overline{c} = \int_0^1 c \cdot p(c)dc=\frac{1}{N}\int_{c^-}^{c^+} \frac{c R(c)}{c(1-c^m)}=\frac{1}{N}\left[I_1(c,m)+\sum_{k=1}^{r} R_k I_{k+1}(c,m) \right]_{c^-}^{c^+}
	\label{eq:cmeanstretch}
\end{equation} 
and
\begin{equation}
	\overline{\omega} = \int_0^1 \omega_m(c) \cdot p(c)dc=\frac{1}{N}\int_{c^-}^{c^+} \frac{\omega_m(c) R(c)}{c(1-c^m)}=\frac{1}{N}\left[c^{m+1}+\sum_{k=1}^{r} R_k\frac{m+1}{m+k+1} c^{m+k+1} \right]_{c^-}^{c^+}
	\label{eq:ommeanstretch}
\end{equation} 
The source term $\omega_x(c)$ in real space can be gained from $\omega(c)$ in $\xi$ space as $\omega_x(c)=(\rho_u s_L)^2\left(\frac{c_p}{\lambda}\right)_u \frac{1}{R(c)} \cdot \omega(c)$, see eq.(\ref{eq:omstrans}), where $R(c)$ takes into account the $c$ variation of $\left(\frac{c_p}{\lambda}\right)$.

\section{Analytical approximation of $\xi(x)$}
We have $c^-=c_m(\xi^-)$ and $c^+=c_m(\xi^+)$ but for non-constant stretch factor the simple relation $\xi^+=\xi^-+\Delta_\xi$ with $\Delta_\xi=r_u\Delta_x$ does not hold any more. It is therefore useful to have an explicit expression $\xi(x)$.
Using the approximation $c(\xi)=c_m(\xi)$ and the polynomial approximation to $R(c)$, one can integrate the differential equation
\begin{equation}
	d\xi/dx=r_u/R(c_m(\xi))
	\label{eq:dxidxR}
\end{equation} 
yielding the implicit equation 
\begin{equation}
	x-x_0=\int_{\xi^-}^{\xi^+} \frac{R(c_m(\xi))d\xi}{r_u}
	\label{eq:xintxiR}
\end{equation} 
Even with the linear approximation to $R(c)$, the integral on the RHS cannot be evaluated analytically. 

\begin{figure} [ht]
	\begin{minipage}[b]{.42\linewidth} % [b] => Ausrichtung an \caption
		\begin{tikzpicture}
			\node[] (Grafik) at (0,0) {\includegraphics[width=1\textwidth]{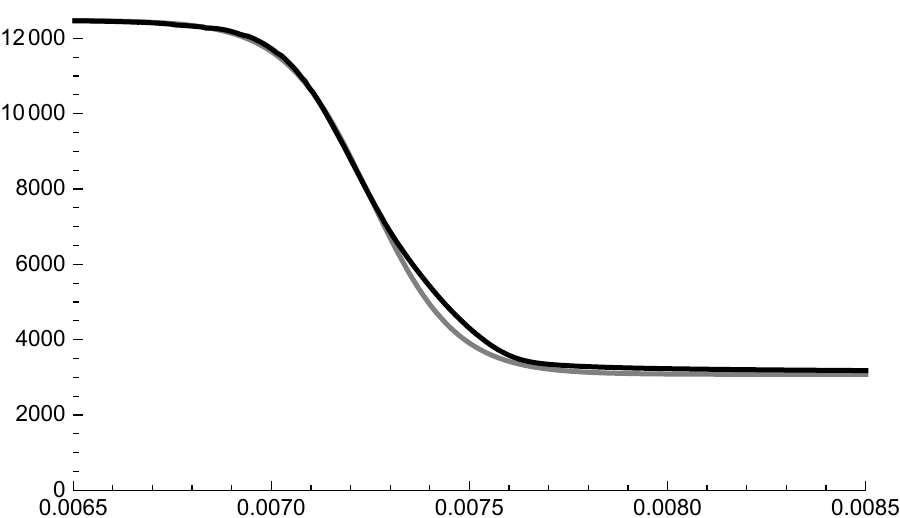}};
			\node[anchor=north,yshift=0pt,xshift=0] at (Grafik.south) {$x$};
			\node[rotate=90,anchor=south,xshift=0pt,yshift=-3] at (Grafik.west) {$d\xi/dx$};
		\end{tikzpicture}
	\end{minipage}
	\hspace{.1\linewidth}% Abstand zwischen Bilder
	\begin{minipage}[b]{.42\linewidth} % [b] => Ausrichtung an \caption
		\begin{tikzpicture}
			\node[] (Grafik) at (0,0) { \includegraphics[width=\linewidth]{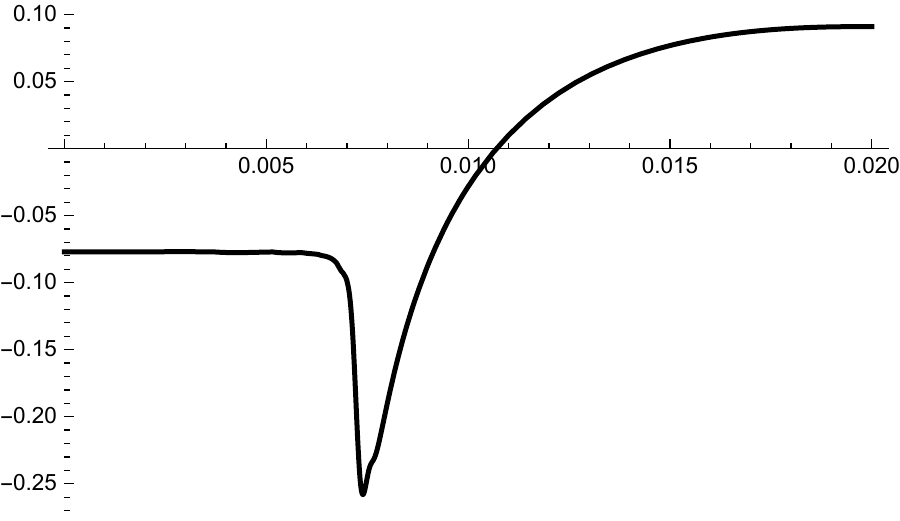}};
			\node[anchor=north,yshift=-0pt,xshift=0] at (Grafik.south) {$x$};
			\node[rotate=90,anchor=south,yshift=-0,xshift=0pt] at (Grafik.west) {$\xi(x)-\xi_a(x)$};
		\end{tikzpicture} 
	\end{minipage}
	\caption{Left:  $d\xi/dx$ evaluated numerically (black), analytic approximation $d\xi_a/dx$ (gray); right: difference between real $\xi(x)$ and approximation $\xi_a(x)$}
	\label{fig:xiapprox}
\end{figure}
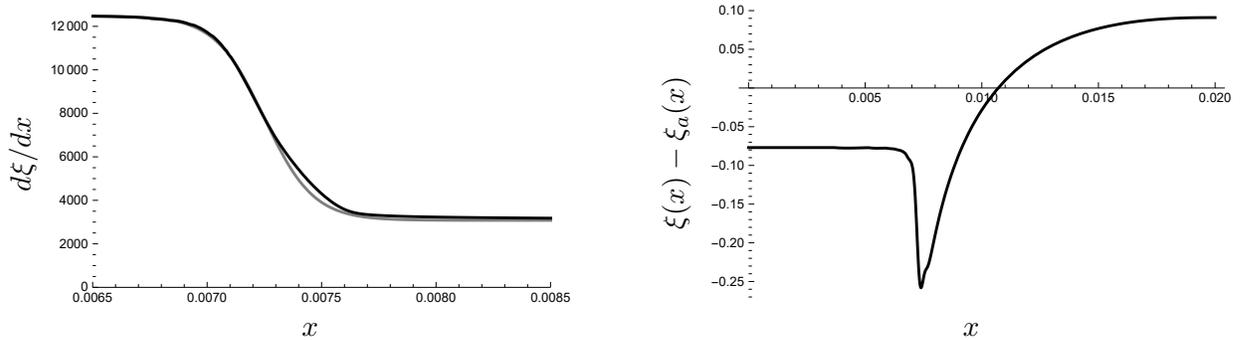

As alternative to numerical interpolation of $\xi(x)$ we provide an approximation to the derivative $d\xi/dx$ directly through a simple ansatz:
\begin{equation}
	r(x)=r_u+\frac{r_b-r_u}{1+exp(-3(r_u-r_b)(x-x_0)/5}
	\label{eq:rxansatz}
\end{equation}
This approximates real $d\xi/dx$ satisfactorily and yields the analytic approximation
\begin{equation}
	\xi_a(x)=\xi_0+\frac{5 (r_b-r_u) \log \left(e^{\frac{3}{5} (r_b+r_u) (x-x_0)}+1\right)}{3 (r_b+r_u)}+r_u x
	\label{eq:xixansatz}
\end{equation}
where $r_b$ is the value of $r(x)$ in the fully burnt region. Fig.(\ref{fig:xiapprox}) shows a comparison of the $d\xi/dx$ profiles and the error in the reproduction of the $\xi(x)$ curve. Fig.(\ref{fig:xidxix}) shows the typical range of $\xi$ values, so the approximation $\xi_a(x)$ is accurate to within $0.2 \%$.

\section{Evaluation of $c^-$,$c^+$ and $N$}
\label{evalc+-}
For application of the laminar flame pdf in a CFD code, the values of $c^-,c^+$ need to be evaluated for a given $\overline{c}$ (which is provided by a $c$ transport equation) and
a given filter $\Delta_x$ in $x$ space. Due to the monotonicity of $c(x)$, $\overline{c}$ will be a monotonic function of $x$ and $\Delta_x$. Similar to \cite{pfitzner2020pdf} 
we propose the following preprocessing steps: \\ \\
- Evaluate $\xi^-=\xi(x)$ and $\xi^+=\xi(x+\Delta_x)$ \\
- Evaluate $c^-=c_m(\xi^-)$ and $c^+=c_m(\xi^+)$ \\
- Evaluate $N$ from eq.(\ref{eq:intpcstretch}) as function of $x,\Delta_x$ \\
- Evaluate $\overline{c}$ from eq.(\ref{eq:cmeanstretch}) using $N$ and $c^-,c^+$ \\
- Evaluate $\overline{\omega}$ from eq.(\ref{eq:ommeanstretch}) using $N$ and $c^-,c^+$ \\ 

$p(c)$ as function of $x, \Delta_x$ is now completely defined, so mean values of other quantities of interest $z(c)$ can be evaluated. Quantities $N,c^-,c^+,$ and $\overline{\omega}$ might be re-tabulated as functions of $\overline{c}$ and filter size $\Delta_x$ for use in a CFD code.

\section{Effects of turbulent flame folding on $p(c)$}
A turbulent flow field at low to moderate Karlovitz numbers will fold the laminar flame front without changing its inner structure noticeably. The simple model with sinusoidal flame folding\cite{pfitzner2020pdf}  showed  that the level of the pdf in the reactive $c$ region will increase by a wrinkling factor $\Xi$ while the pdf near $c^-,c^+$ is smeared out due to pushing in/out of parts of isosurfaces from the filter volume by flame folding. The mean reaction rate $\overline{\omega(c)}$ will mainly increase through a wrinkling factor $\Xi$. A similar behaviour was observed from analysis of DNS data\cite{tsui2014linear}$^,$\cite{moureau2011large}$^,$\cite{lapointe2017priori}. At higher Karlovitz number into the corrugated / thin reaction zone regime, a stronger mixing of the  preheat zone will occur through small turbulent eddies. 

The exact relationships\cite{pfitzner2020pdf} evaluated in $x$ space provide:
\begin{equation}
	p(c) = \frac{1}{\Omega_x}  \int_{\Omega_x} \delta (c (\vec{x}) -
	c) d \vec{x} 
	\label{eq:pcintx}
\end{equation}
The $c$ isosurface area density within the volume $\Omega_x$ is given by
\begin{equation}
	\sigma (c) = \frac{1}{\Omega_x} \int_{\Omega_x} \delta (c (\vec{x}) - c) \mid
	\nabla c (\vec{x}) \mid d \vec{x} 
	\label{eq:Sigma}
\end{equation}
and a correction factor $I(c)$ can be defined as
\begin{equation}
	I(c)=\frac{\frac{1}{\Omega_x} \int_{\Omega_x} \delta (c (\vec{x}) - c) \mid
		dc / dx \mid_{1D,c} d \vec{x}}{\sigma(c)}
	\label{eq:Idef}
\end{equation}
yielding
\begin{equation}
	p(c) =  \frac{\sigma(c) \cdot I(c)}{\mid dc / dx \mid_{1D,c}}
	\label{eq:pcdecomp}
\end{equation}
where $\Omega_x$ is the filter volume in real space. $I(c)$ represents the effect of the mean difference between the local $c$ gradient and the one of the  1-D flat flame.

With a nonlinear transformation, eq.(\ref{eq:pcdecomp}) becomes 
\begin{equation}
	p(c) =  \frac{\sigma(c) \cdot I(c) \cdot R(c)}{r_u \cdot \mid dc / d\xi \mid_{1D,c}}
	\label{eq:pctrans}
\end{equation}

Analysis of DNS data of turbulent plane premixed flames at $u'/s_L=5,15$\cite{pfitzner2021pdf} have shown that despite considerable flame
folding, $I(c)$ and $\sigma(c)$ are rather constant in $c$ for large, RANS-type filter volumes. $I(c)$ was near one and the level of $\sigma(c)$ raised proportional to flame wrinkling intensity. The role of the cutoffs $c^-,c^+$ of the 1D case is taken into account through the $c$ dependence of $\sigma(c)$ in a three-dimensional (3D) setting. $\sigma(c)$ then drops continuously to zero near $c^-$,$c^+$.

The pdf of the reaction layer in a thin folded flame can be approximated by the 1D laminar flame pdf through use of a reduced filter width $\Delta'=\Delta/\Xi$, where $\Xi$ is a wrinkling factor:
\begin{equation}
	p_\Xi(c,\Delta) = p_{1D, \Xi=1}(c,\Delta'=\Delta/\Xi)
	\label{eq:pcfolded}
\end{equation}
where also $c^-,c^+$ are evaluated with a filter scaled by $1/\Xi$. This will automatically rise the
level of the pdf in the range of $c^-,c^+$ and the evaluation of $c^-,c^+$ using $\Delta'=\Delta/\Xi$ will ensure the correct normalization of $p(c)$.

\section{PDF of partially premixed flames}
In the case of thin flame fronts propagating through a fuel-air mixture with locally (slowly) varying mixture fraction $Z$, a common model for the joint  pdf $p(Z,c)$ is based on application of Bayes's theorem\cite{dovizio2016}:
\begin{equation}
	p(Z,c) = p_Z(Z) \cdot p(c \mid Z)
	\label{eq:pZc}
\end{equation}
where $p_Z(Z)$ is the marginal pdf of mixture fraction and $p(c \mid Z)$ is the pdf of $c$ conditional on mixture fraction, which is modelled as 1D laminar premixed flame pdf at $\phi(Z)=\frac{Z}{1-Z} \cdot \frac{Z_{st}}{1-Z_{st}}$ corresponding to the respective mixture fraction. This assumes that the inner structure of the propagating flame fronts are little affected by the spatial variation of $Z$.

$p_Z(Z)$ is commonly modelled as a $\beta$ function:
\begin{equation}
	p_Z(Z)=p_\beta(Z)=\frac{Z^{a-1}(1-Z)^{b-1}\Gamma(a+b)}{\Gamma(a)\Gamma(b)} 
	\label{eq:pZbeta}
\end{equation}
with parameters $a,b$ evaluated from a transport equations of mixture fraction $Z$ and its variance. In LES, the latter is sometimes also evaluated from an algebraic model.

The conditional pdf $p(c \mid Z)$ modelled as laminar premixed flame pdf is then given by
\begin{equation}
	p(c \mid Z)= \frac{1}{N_Z}\frac{R_Z(c)}{dc_{m_Z}/d\xi}H(c-c_Z^-)H(c_Z^+-c)
	\label{eq:pccond1D}
\end{equation}
with $m(Z),R_Z(c)$ and $c^-,c^+$ evaluated at the $\phi(Z)$. In the 3D case, the conditional pdf would read
\begin{equation}
	p(c \mid Z)=  \frac{\sigma_Z(c) \cdot I_Z(c) \cdot R_Z(c)}{\mid dc_{m_Z} / d\xi \mid_{1D,c}}
	\label{eq:pccond3D}
\end{equation}
Again, for low to moderate Karlovitz number one can expect $I_Z(c) \approx 1$ in the
reactive $c$ region. $\sigma(c)$ will be dominated by wrinkling of large turbulent eddies, its cutoff regions will depend on the ratio of filter volume $\Omega_x$ to laminar flame thickness $\delta_{th}(Z)$. Since the latter grows as $\phi$ moves away from $\phi=1$, a smaller subgrid isosurface density can be expected in lean regions compared to near-stoichiometric ones. Since the maximum of $c_p/\lambda$ varies only by $30\%$ over the investigated $\phi$ range (mainly in the lean region), the stretch factor is dominated by the $\phi$ variation of $s_L$. Inspection of fig.(\ref{fig:tmaxsl}) shows that a flame at $\phi=0.5$ is six times wider than a stoichiometric flame.

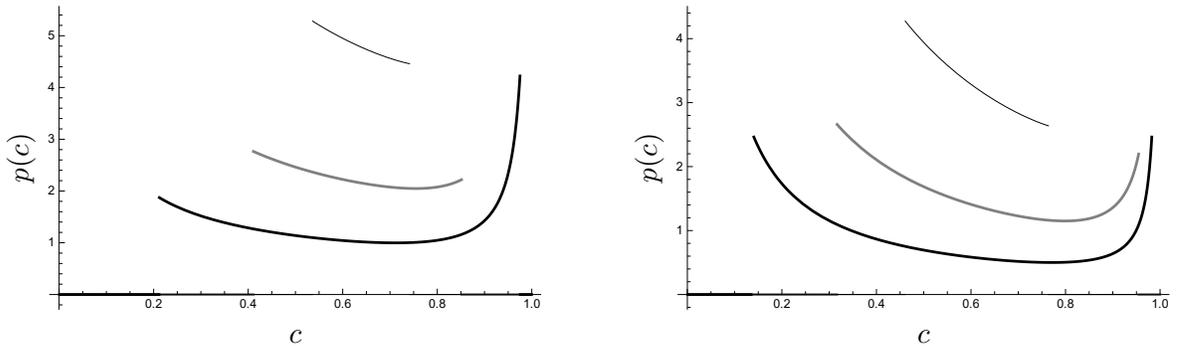
\begin{figure} [ht]
	\begin{minipage}[b]{.4\linewidth} % [b] => Ausrichtung an \caption
		\begin{tikzpicture}
			\node[] (Grafik) at (0,0) {\includegraphics[width=1\textwidth]{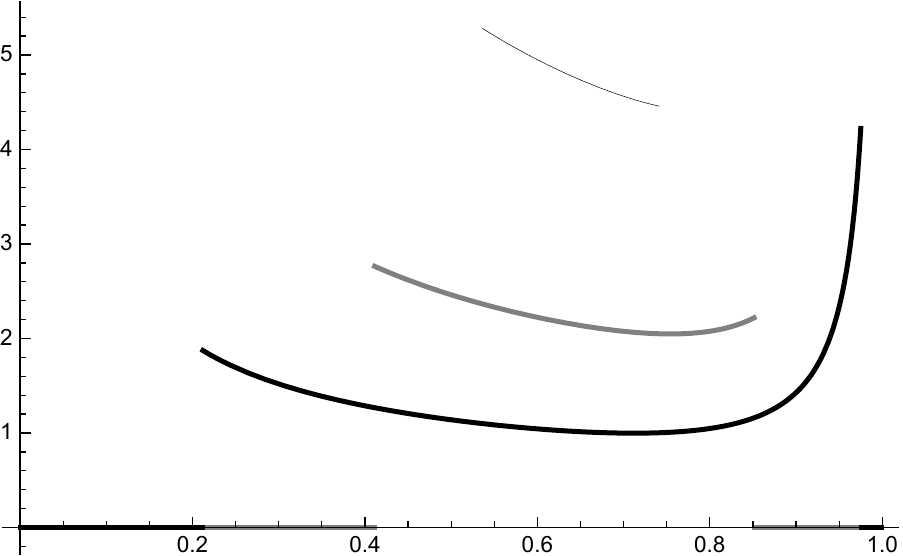}};
			\node[anchor=north,yshift=0pt,xshift=0] at (Grafik.south) {$c$};
			\node[rotate=90,anchor=south,xshift=0pt,yshift=-3] at (Grafik.west) {$p(c)$};
		\end{tikzpicture}
	\end{minipage}
	\hspace{.1\linewidth}% Abstand zwischen Bilder
	\begin{minipage}[b]{.4\linewidth} % [b] => Ausrichtung an \caption
		\begin{tikzpicture}
			\node[] (Grafik) at (0,0) {\includegraphics[width=1\textwidth]{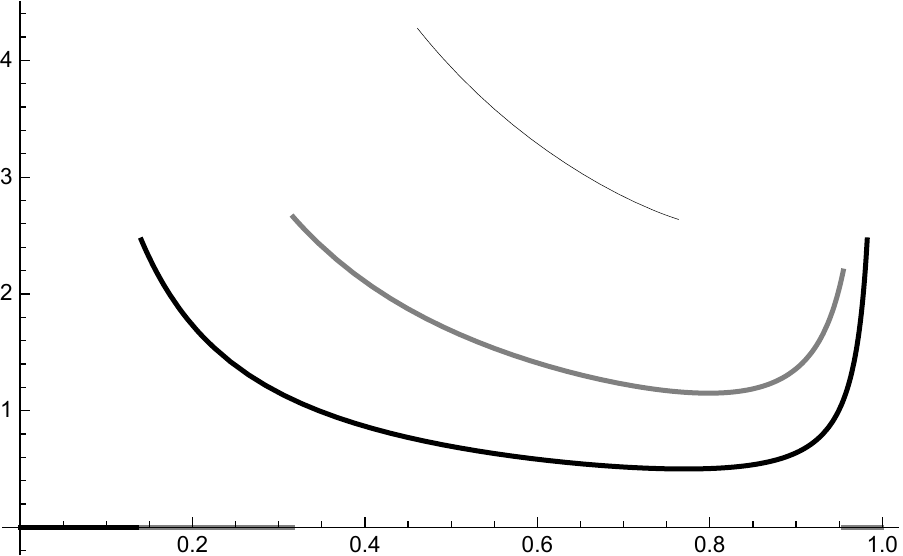}};
			\node[anchor=north,yshift=0pt,xshift=0] at (Grafik.south) {$c$};
			\node[rotate=90,anchor=south,xshift=0pt,yshift=-3] at (Grafik.west) {$p(c)$};
		\end{tikzpicture}
	\end{minipage}
	\caption{$p(c)$ for $\overline{c}=0.6$,  $\Delta_x=240 \mu$ and $\phi=1$ (Black), $\phi=0.6$ (Gray),$\phi=0.5$ (Black, thin); left: full $c_p/\lambda$, right:$c_p/\lambda$=$(c_p/\lambda)_u$ }
	\label{fig:pcphisel}
\end{figure}

A certain filter interval $\Delta_x$ in $x$ space then translates into a smaller $\Delta_\xi$ in $\xi$ space  and $c^-,c^+$ are nearer to $\overline{c}$ for lean/rich flames than for stoichiometric ones. Fig.(\ref{fig:pcphisel}) shows 1D unstrained flame pdfs at $\phi=1,0.6,0.5$ in a cell of width $\Delta_x=240 \mu$ and for $\overline{c}=0.6$ using the full nonlinear $R(c)$ (left) and using $R(c)=1$ (right), i.e. performing the coordinate transformation using the unburnt $c_p/\lambda$ only. The pdf's for the leaner flames cover a much smaller range $[c^-,c^+]$ in transformed $\xi$ coordinates than the stoichiometric one. Note however that application of a wrinkling factor $\Xi$ will raise the pdf level and narrow its range, so a partial correction of the difference can occur.

\begin{figure} [ht]
	\begin{minipage}[b]{.46\linewidth} % [b] => Ausrichtung an \caption
		\begin{tikzpicture}
			\node[] (Grafik) at (0,0) {\includegraphics[width=1\textwidth]{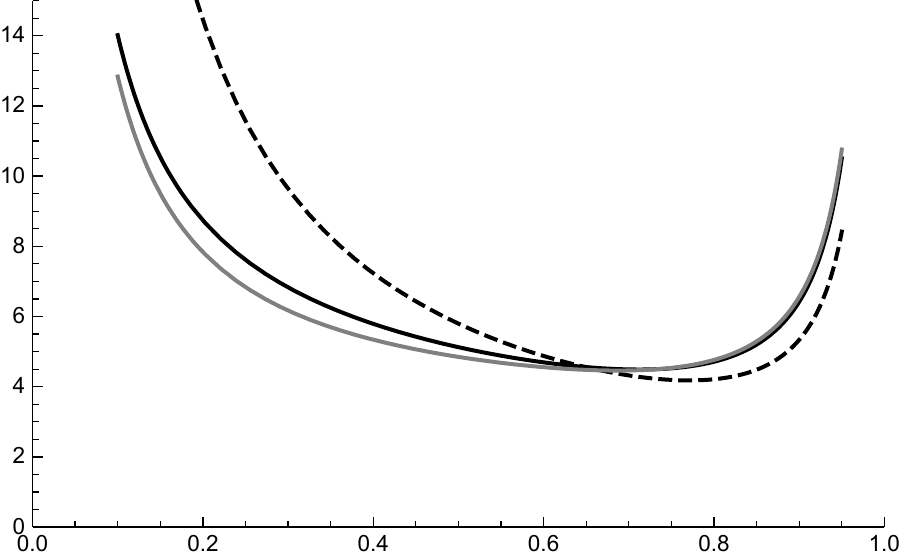}};
			\node[anchor=north,yshift=0pt,xshift=0] at (Grafik.south) {$c$};
			\node[rotate=90,anchor=south,xshift=0pt,yshift=-3] at (Grafik.west) {$F(c)/(dc/d\xi)$};
		\end{tikzpicture}
	\end{minipage}
	\caption{$R(c)/(dc/d\xi)$; black: full nonlinear $R(c)$, dashed: $R(c)=1+\frac{2(R(1)-1)}{3}$= const., gray: $R(c)=1+ \frac{7r_u}{10r_b}c$}
	\label{fig:pdfnonlinlin}
\end{figure}

Fig.(\ref{fig:pdfnonlinlin}) shows a comparison of the shape of the non-normalized pdf $R(c)/(dc/d\xi)$, scaled with the full polynomial $R(c)$, using a constant value $R(c)=1+\frac{2}{3}(R(1)-1)$ and using the linear approximation $R(c)=1+ \frac{7r_u}{10r_b}c$. The shape of the pdf cannot be recovered using a constant $R$, while a quite good approximation is possible with the linear approximation to $R(c)$. 

\section{Analytical partially premixed flame pdf for constant $c_p/\lambda$}
A fully explicit analytic pdf $p(Z,c)$ can be formulated in 
the case that $c_p/\lambda$ depends only on mixture fraction $Z$, but does not change across the flame. 
We present this here to demonstrate the main ingredients of the pdf and because this form can be applied directly in the evaluation of DNS data which use this assumption. 

Assuming a beta pdf for marginal pdf of mixture fraction and a laminar flame pdf for $p(c \mid Z)$ filtered with $\Delta'=\Delta/\Xi$, we obtain
\begin{equation} 
    p(Z,c)=\frac{Z^{a-1}(1-Z)^{b-1}\Gamma(a+b)}{\Gamma(a)\Gamma(b)} \cdot \frac{\Xi}{\rho_u s_L(Z) (c_p / \lambda)_u \cdot \Delta_x} \cdot \frac{1}{c(1-c^{m(Z)})}H(c-c^-)H(c^+-c)
	\label{eq:pdfpartpre}
\end{equation}
where parameters $a,b$ of the beta pdf for $Z$ are evaluated from the $Z$ mean and variance, $s_L(Z)$ and $m(Z)$ are preprocessed functions of $Z$ as indicated above and $c^-,c^+$ are calculated similarly to the purely premixed case from  $\overline{(c \mid Z)}$ and the effective filter width $\Delta'_\xi=\rho_u s_L(Z) (c_p / \lambda)_u \cdot \Delta_x / \Xi$. The $c^-,c^+$ correlation provided in \cite{pfitzner2020pdf} might be used as an approximation. 
The wrinkling factor $\Xi$ might be evaluated from models developed in the framework of ATF or FSD models or from a
transport equation of the $c$ variance, using the FLF pdf method to evaluate $\Delta'=\Delta/\Xi$.
Also here it should be considered that the subgrid flame wrinkling factor is expected to increase with increasing $\Delta/\delta_{th}$, so $\Xi$ should be higher in regions $\phi \approx 1$ than in very lean or very rich regions with larger $\delta_{th}$.

For small, LES-typical filter widths, the smearing of the 3D pdf near $c^-,c^+$ is noticeable, see pdfs shown in\cite{moureau2011large}$^,$\cite{lapointe2017priori}$^,$\cite{pfitzner2020pdf}. For larger, RANS-like filter volumes, the effect is minor since $c^-,c^+$ are very near to $c=0,1$ and the smear out of the pdf in this region has only a small effect on most mean values.

\section{Conclusions and outlook}
In this contribution we have shown that the progress variable distribution of the  $CO_2+CO$ and $H_2O+H_2$   specific mole number combinations in free premixed laminar flame profiles generated with the GRI mech 3.0 detailed chemistry mechanism using real transport coefficients can be represented very accurately by a slight generalization of recently proposed analytical flame profiles after a canonical stretch of coordinates. 

The model parameter $m$ of the analytical profiles in transformed coordinates can be calculated very accurately using the laminar flame speed and the burnt temperature only, using results from single-step Arrhenius chemistry theory. Results indicate that slightly different effective activation temperatures for the Arrhenius model are required to reproduce flame profiles at different fuel/air ratios $\phi$.

Profiles of strained flames can be also be reproduced accurately after application of an additional linear coordinate transformation. The stretch parameter $a$ is found to be a linear function of a Karlovitz number. As expected from combustion theory, the activation temperature and the thermal flame thickness are approximately independent of strain level in the investigated range of $K$.

The analyses permit the derivation of an analytical presumed laminar flame pdf for detailed chemistry premixed methane flames (Le=1) at different fuel/air ratio. The effect of a nonlinear $c_p/\lambda$ on the pdf is derived. $\overline{c}$, the $c$ variance and $\overline{\omega}$ can be evaluated analytically also this case. We also provide an analytic approximation of $\xi(x)$.

We discuss modifications of the pdf due to turbulence shortly and propose an analytical pdf $p(Z,c)$  based on an application of Bayes' theorem for the case of propagation of thin flames into non-homogeneous mixture with slowly varying mixture fraction. The marginal pdf of mixture fraction is approximated by a beta function, while the pdf of $c$ conditional on $Z$ is represented by pdf's of premixed laminar flame profiles at $\phi$ corresponding to the local $Z$. The slow variation of $Z$ translates into the assumption of negligible effects from cross-diffusion of mixture fraction between different $c$ profiles, i.e. $\mid \nabla c \mid \gg \mid \nabla  Z \mid$ in the reaction region. This is a good assumption in many applications featuring such flames.

The next steps will be the validation of the proposed detailed chemistry pdfs using DNS data of fully premixed flames generated with the same detailed chemistry mechanism and transport. The partially premixed analytic pdf will be first validated with DNS databases of flame propagating into non-homogeneous mixture using single-step Arrhenius chemistry. The final step will be the validation of the pdf of partially premixed flames with detailed chemistry using
suitable DNS databases. 

Further investigations are planned to derive / validate models of the wrinkling factor $\Xi$ and the subgrid flame surface density $\sigma(c)$ and, if necessary, the correction factor $I(c)$. $\Xi$ models developed in the framework of flame surface density and ATF-type models will be the starting point. 

We will also investigate the suitability of models of the subgrid variance, which showed some promise to determine the wrinkling factor using the FLF pdf method. A posteriori simulations will be performed to on DNS and experimental configurations. Extensions of the approach to cases with more the one progress variable will be sought.

\section{Acknowledgements}
The authors would link to thank M. Klein and N. Chakraborty for fruitful discussions. We also gratefully acknowledge funding of part of this work through Deutsche Forschungsgemeinschaft in projects PF443/9-1 and KL1456/5-1.
\section{Author's contributions}
F. Breda generated the GRI mech 3.0 premixed flame profiles and data. M. Pfitzner performed all other analyses.
\section{Data availability statement} The data that support the findings of this study are available from the corresponding author upon reasonable request.
\section{Compliance with Ethical standards} The authors confirm that Ethical standards have been obeyed. 
\section{Conflict of interest} The authors declare that they have no conflict of interest.

% If in two-column mode, this environment will change to single-column format so that long equations can be displayed. 
%\begin{widetext}
%$$\mbox{put long equation here}$$
%\end{widetext}

% Figures should be put into the text as floats. 
% Use the graphics or graphicx packages (distributed with LaTeX2e).
% Put the label that you will use with \ref{} command in the braces of the \label{} command.
%
% \begin{figure}
% \end{figure}

% \begin{table}
% \end{tabular}
% \end{table}

%\begin{acknowledgments}
% Put your acknowledgments here.
%\end{acknowledgments}

\bibliography{POF_OBrien_Pfitzner.bib}

\end{document}